\documentclass[sn-aps, referee, pdflatex]{sn-jnl}
\usepackage[T1]{fontenc}

\usepackage{graphicx}
\usepackage{multirow}
\usepackage{amsmath,amssymb,amsfonts}
\usepackage{amsthm}
\usepackage{mathrsfs}
\usepackage[title]{appendix}
\usepackage{xcolor}
\usepackage{textcomp}
\usepackage{manyfoot}
\usepackage{algorithm}
\usepackage{algorithmicx}
\usepackage{algpseudocode}
\usepackage{listings}
\usepackage{todonotes}
\usepackage[version=4]{mhchem}
\usepackage{siunitx}
\usepackage{hyperref}
\usepackage[subrefformat=parens]{subcaption}
\usepackage{geometry}
\usepackage{ulem}
\usepackage{threeparttable}
\usepackage{booktabs}
\usepackage{xr}
\usepackage{physics}
\usepackage{multirow}
\let\orcidlogo\relax
\usepackage{orcidlink}
\usepackage{datetime}

\makeatletter
\newcommand*{\addFileDependency}[1]{
  \typeout{(#1)}
  \@addtofilelist{#1}
  \IfFileExists{#1}{}{\typeout{No file #1.}}
}
\makeatother
\newcommand*{\arxivexternaldocument}[2][]{
  \externaldocument[#1]{#2}
  \addFileDependency{#2.aux}
}

\arxivexternaldocument[si-]{xref/si}

\theoremstyle{thmstyleone}

\theoremstyle{thmstyletwo}

\theoremstyle{thmstylethree}

\newdateformat{monthyeardate}{
  \monthname[\THEMONTH], \THEYEAR}

\begin{document}

\title{\LARGE{\textbf{MatterSim-MT: A multi-task foundation model for \textit{in silico} materials characterization}}}

\author*[2]{\fnm{Han} \sur{Yang}\orcidlink{0000-0002-4531-093X}}\email{yanghan234@outlook.com}
\equalcont{These authors contributed equally to this work.}

\author[2]{\fnm{Xixian} \sur{Liu}\orcidlink{0009-0008-9215-3990}}
\equalcont{These authors contributed equally to this work.}

\author[2]{\fnm{Chenxi} \sur{Hu}\orcidlink{0009-0006-8486-9230}}
\equalcont{These authors contributed equally to this work.}

\author[2]{\fnm{Yichi} \sur{Zhou}}
\equalcont{These authors contributed equally to this work.}

\author[2]{\fnm{Yu} \sur{Shi}\orcidlink{0000-0001-9235-8963}}
\equalcont{These authors contributed equally to this work.}

\author[2]{\fnm{Chang} \sur{Liu}\orcidlink{0000-0001-5207-5440}}
\equalcont{These authors contributed equally to this work.}

\author[2]{\fnm{Junfu} \sur{Tan}\orcidlink{}}
\equalcont{These authors contributed equally to this work.}

\author*[2]{\fnm{Jielan} \sur{Li}\orcidlink{0000-0003-4428-2452}}\email{jielanlee@outlook.com}
\equalcont{These authors contributed equally to this work.}

\author[2]{\fnm{Guanzhi} \sur{Li}\orcidlink{0000-0002-4167-6432}}
\equalcont{These authors contributed equally to this work.}

\author[2]{\fnm{Qian} \sur{Wang}\orcidlink{0009-0007-7680-4514}}
\equalcont{These authors contributed equally to this work.}

\author[2]{\fnm{Yu} \sur{Zhu}\orcidlink{0009-0007-1068-4548}}
\equalcont{These authors contributed equally to this work.}

\author[2]{\fnm{Zekun} \sur{Chen}\orcidlink{0000-0002-4183-2941}}
\equalcont{These authors contributed equally to this work.}

\author[2]{\fnm{Shuizhou} \sur{Chen}\orcidlink{0009-0005-2701-5565}}
\equalcont{These authors contributed equally to this work.}

\author[1]{\fnm{Fabian L.} \sur{Thiemann}\orcidlink{0000-0003-2951-6740
}}

\author[1]{\fnm{Claudio} \sur{Zeni}\orcidlink{0000-0002-6334-2679}}
\author[2]{\fnm{Matthew} \sur{Horton}\orcidlink{0000-0001-7777-8871}}
\author[1]{\fnm{Robert} \sur{Pinsler}\orcidlink{0000-0003-1454-188X}}
\author[1]{\fnm{Andrew} \sur{Fowler}\orcidlink{0000-0002-7360-3078}}
\author[1]{\fnm{Daniel} \sur{Z\"ugner}\orcidlink{0000-0003-1626-5065}}
\author[2]{\fnm{Tian} \sur{Xie}\orcidlink{0000-0002-0987-4666}}
\author[1]{\fnm{Lixin} \sur{Sun}\orcidlink{0000-0002-7971-5222}}
\author[2]{\fnm{Yicheng} \sur{Chen}\orcidlink{0009-0002-3829-4473}}
\author[2]{\fnm{Lingyu} \sur{Kong}\orcidlink{0009-0006-2226-5730}}
\author[1]{\fnm{Yeqi} \sur{Bai}\orcidlink{0000-0001-5959-3435}}
\author[1]{\fnm{Deniz} \sur{Gunceler}\orcidlink{0009-0000-6654-6856}}
\author[1]{\fnm{Frank} \sur{No\'e}\orcidlink{0000-0003-4169-9324}}
\author*[2]{\fnm{Hongxia} \sur{Hao}\orcidlink{0000-0002-4382-200X}}\email{hongxiahao19@gmail.com}
\author*[2]{\fnm{Ziheng} \sur{Lu}\orcidlink{0000-0003-2239-8526}}\email{zluag@connect.ust.hk}

 \affil[1]{\orgname{Microsoft Research AI for Science}}
 \affil[2]{\orgname{This work was carried out while the author was affiliated with Microsoft Research}}

\abstract{
Accurate property characterization is a major bottleneck in materials design.
While first‑principles methods and task‑specific machine‑learning models have driven important progress, they remain fundamentally limited in scalability and generalizability across the vast space of structures and properties relevant to real‑world materials design.
We present \texttt{MatterSim-MT}, a multi-task foundation model for \textit{in silico} materials simulation and property characterization.
The model is pretrained on over 35 million first-principles-labeled structures covering 89 elements, temperatures up to \SI{5000}{\kelvin} and pressures up to \SI{1000}{\giga\pascal}, and is fine-tuned on various properties including Bader charges, magnetic moments, Born effective charges, and dielectric matrices. Out of the box, \texttt{MatterSim-MT} not only serves as a foundation model for predicting material structure, dynamics and thermodynamics, its multi-task architecture also enables a wide range of complex simulations that cannot be captured by potential energy surfaces alone. For example, we demonstrate pressure-dependent LO-TO phonon splitting in \ce{SiC} with close agreement with experiment, electric hysteresis in ferroelectric \ce{BaTiO3}, and the cationic-to-anionic redox transition during delithiation of a Li-rich cathode material.
Finally, we show that \texttt{MatterSim-MT} scales well with more data and parameters, can be efficiently fine-tuned to higher levels of theory, and can be efficiently  extended to new systems via active learning. Overall, we believe this approach provides a scalable route to accurate \textit{in silico} materials characterization.
}

\maketitle

\newpage

\section{Introduction}

Materials design drives a wide range of technological advances, for example in nanoelectronics,\cite{fiori2014electronics,li2007electronic} energy storage,\cite{mizushima1980lixcoo2,ceder1998identification} biomedicine,\cite{tibbitt2015progress} and environmental sustainability\cite{li2018review,hu2010design}. However, the enormous space of potential candidates makes it challenging to find materials with the required properties under desired operating conditions. This problem is further exacerbated by the slow pace of traditional experimental synthesis and characterization.
Recent advances in large-scale materials databases\cite{hellenbrandt2004inorganic,jain2013commentary,curtarolo2013high,kirklin2015open,draxl2019nomad,schmidt2022large,schmidt2022dataset,merchant2023scaling,barroso2024open,horton2025accelerated,cavignac2025alexandria} and generative models for materials\cite{zeni2025generative,yang2023scalable,yang2024generative,miller2024flowmm,cao2025crystalformer,sriram2024flowllm} have accelerated the generation of hypothetical structures with desired target properties. Yet, the ability to rapidly generate candidates has only increased the burden on experimental synthesis and property characterization.

Universal machine learning interatomic potentials (uMLIPs) aim to accelerate the materials design process by providing accurate stability and property predictions for a wide range of materials, improving candidate selection. Inspired by the success of foundation models in areas such as natural language processing\cite{brown2020language,touvron2023llama,chowdhery2022palm,achiam2023gpt,grattafiori2024llama,geminiteam2024gemini,guo2025deepseek} and image generation,\cite{ramesh2022hierarchical,rombach2022high} a growing number of deep-learning-based uMLIPs trained on large-scale atomistic data for materials have been proposed.\cite{chen2022universal,batatia2023foundation,neumann2024orb,rhodes2025orb,deng2023chgnet,zhang2023dpa,merchant2023scaling,barroso2024open,xie2024gptff,shoghi2023molecules,choudhary2021atomistic,park2024scalable,liao2024equiformerv2,mazitov2025petmad,yang2024mattersim}
Similar to foundation models in other fields, atomistic foundation models should support a wide range of downstream tasks.
Yet, the dominant paradigm is to train uMLIPs exclusively on potential energy surface (PES) data.\cite{batatia2023foundation,chen2022universal,neumann2024orb,rhodes2025orb} While such uMLIPs accelerate structure optimization and molecular dynamics simulations,\cite{unke2021machine} many materials problems---such as predicting the dielectric response or polarization\cite{resta1994macroscopic,baroni2001phonons}---cannot be solved by computing energies, forces, or stresses alone.
To address this, a few efforts have sought to move beyond MLIPs by predicting additional material properties.\cite{falletta2025unified,deng2023chgnet,zhou2026matris,zhong2025machine,martin2025general} However, these approaches are often limited to expert models tailored to specific systems and tasks, or evaluated only on machine learning benchmarks rather than realistic simulation workflows.

We present \texttt{MatterSim-MT}, a multi-task (MT) foundation model for \textit{in silico} materials simulation and property characterization. The model learns a unified atomic representation trained to jointly predict multiple physical properties derived from first-principles calculations across a large range of materials.
By leveraging active learning and integrating molecular dynamics simulations with large-scale first-principles computations,\cite{hohenberg1964inhomogeneous,kohn1965self,ganose2025atomate2} we construct a dataset of atomic structures covering 89 elements, temperatures up to \SI{5000}{\kelvin}, and pressures up to \SI{1000}{\giga\pascal}. We pretrain \texttt{MatterSim-MT} on this data to enable first-principles-quality characterization of material stability, phonon spectra, mechanical and transport properties under wide temperature and pressure ranges. For example, \texttt{MatterSim-MT} predicts maximum phonon frequencies with \SI{1}{\tera\hertz} error and provides more accurate predictions of phase stability for binary phase diagrams than previous methods, especially at higher temperatures and pressures.
We further fine-tune the model on existing first-principles calculations of various materials properties to enable accurate property prediction. Beyond energies, forces and stresses,  \texttt{MatterSim-MT} natively supports predictions for magnetic moments, Bader charges, Born effective charges (BEC), and dielectric matrices. For instance, the model achieves Born effective charges with 0.08 e error on inorganic solids relative to first-principles calculations. More importantly, our multi-task architecture enables complex simulations of materials beyond the PES using a single unified model, including the simulation of materials under external conditions such as electrical fields. As a demonstration,
we reproduce pressure-dependent LO-TO phonon splitting in SiC with close agreement
with experiment, electric hysteresis in ferroelectric BaTiO3, and the cationic-to-anionic redox
transition during delithiation of a Li-rich cathode material. Finally, we show that \texttt{MatterSim-MT} predictably improves with more data and model parameters, can be efficiently fine-tuned to higher levels of theory, and can be efficiently extended to new systems via active learning.
Together, these capabilities establish \texttt{MatterSim-MT} as a versatile foundation model for materials simulation and property characterization across a wide range of operating conditions.

\section{Learning the materials space}

\begin{figure}
    \centering
    \includegraphics[width=\textwidth]{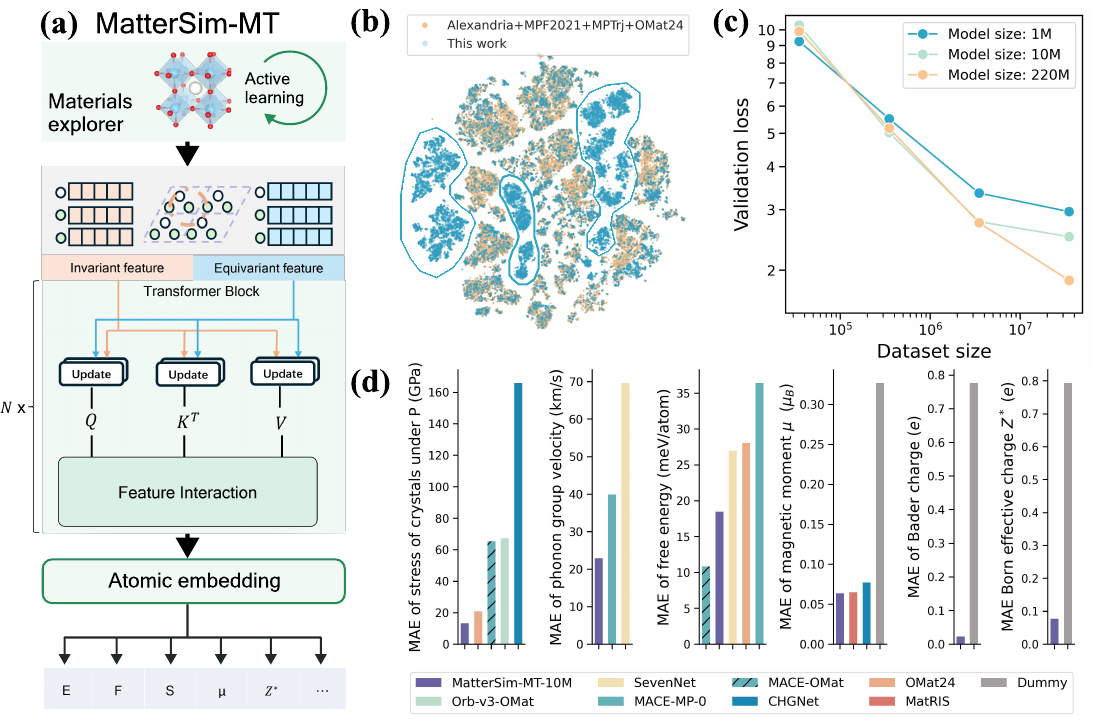}
    \caption{\textbf{MatterSim-MT is a scalable materials foundation model trained on a large, diverse dataset, achieving state-of-the art performance on a variety of prediction tasks}.
    \textbf{(a)} Illustration of \texttt{Mattersim-MT} architecture: The dataset is enriched by an active-learning-based materials explorer (top), which is used to train a transformer architecture with invariant and equivariant features (center), providing an atomic embedding from which various materials properties are predicted (bottom);
    \textbf{(b)} t-SNE analysis of the atomic embedding of the first 20 elements selected from the dataset used in this work compared with the combination of MPF2021, MPtraj, Alexandria and OMat24 datasets;
    \textbf{(c)} Validation loss of energy prediction with \texttt{MatterSim-MT} given different model and dataset sizes;
    \textbf{(d)} Performance comparison of \texttt{MatterSim-MT-10M}, OMat24 (\texttt{eqV2-M OMat MPtrj-sAlex}),\cite{barroso2024open} MACE-MP-0,\cite{batatia2023foundation}, MACE-OMat, CHGNet,\cite{deng2023chgnet} Orb-v3,\cite{rhodes2025orb} SevenNet,\cite{park2024scalable}, MatRIS,\cite{zhou2026matris}, and a dummy predictor that predicts the mean value of the dataset.
    }
    \label{fig:overview}
\end{figure}

\texttt{MatterSim-MT} learns universal representations of materials across a wide range of temperatures and pressures by pretraining on large-scale atomistic data to reproduce a unified first-principles PES. The model is further fine-tuned on additional properties including magnetic moments ($\mu$; assuming ferromagnetic ordering), Bader charges, dielectric matrices ($\varepsilon_\infty$), and Born effective charges ($Z^*$) (see \autoref{si-sec:training-data}). Together, these tasks ensure that the model learns a unified representation of atomistic systems for predicting a wide range of materials properties, including structures, energetics, dynamics and thermodynamics, while also supporting multi-task simulations that requires prediction of properties beyond potential energy surfaces.
This makes it possible to transfer knowledge across materials classes, properties, and operating conditions at scale.

\texttt{MatterSim-MT} builds on the iterative active learning framework introduced in Ref.~\citenum{yang2024mattersim} to generate a large, diverse first-principles dataset spanning ground-state and off-equilibrium structures over wide temperature and pressure ranges (\autoref{fig:overview}(a), top). Underexplored structures are selectively labeled with first-principles calculations and used to continue improving the model. This process yields a dataset with broad chemical and configurational coverage, comprising \SI{35}{M} atomic structures and approximately \SI{450}{M} atomic force vector labels.
To illustrate the diversity of the sampled configurational space, \autoref{fig:overview}(b) shows a t-SNE analysis of atomic-level embeddings for the first 20 elements (see \autoref{si-fig:tsne-89elements} for an analysis across all 89 elements). Compared to datasets constructed from near-equilibrium relaxations\cite{deng2023chgnet,schmidt2022dataset,chen2022universal,jain2013commentary} or short \textit{ab initio} molecular dynamics simulations\cite{barroso2024open}, the dataset used in this work exhibits broader configurational coverage (see circled regions in \autoref{fig:overview}(b)). Compared to \texttt{MatterSim-v1} in Ref.~\citenum{yang2024mattersim}, the dataset size increases from \SI{6}{M} to \SI{35}{M} structures and covers structures under a more continuous range of temperatures of \SIrange{0}{5000}{\kelvin} and pressures of \SIrange{0}{1000}{\giga\pascal}.
This extension facilitates a wide range of downstream tasks from materials synthesis and reactions to high-pressure applications (see \autoref{si-fig:tp-distribution-of-public-dataset}).

The model architecture of \texttt{MatterSim-MT} is based on GeoMFormer\cite{chen2023geomformer}, a scalable, transformer-based architecture for atomistic simulation. We further incorporate  physical priors to enforce energy conservation, respect periodic boundary conditions, and preserve translational and rotational invariance (\autoref{fig:overview}(a); see \autoref{si-sec:model-arch} for details).
In addition, we ensure that the model is able to make effective use of as much data as possible given that first-principles labels are expensive to generate. In \autoref{fig:overview}(c), we show that the validation loss of \texttt{MatterSim-MT} keeps decreasing as the dataset size and model size increase. In practice, scaling the model size is limited by the speed requirements of the downstream task. We find that a \SI{10}{M} parameter model strikes a balance between accuracy and throughput for computationally intensive tasks such as running molecular dynamics simulations.
In the following, we use \texttt{MatterSim-MT-10M} with \SI{10}{M} parameters for predictions and simulations.

\section{MatterSim-MT as a machine learning interatomic potential}
\label{sec:foundation_model}

\begin{figure}
    \centering
    \includegraphics[width=\linewidth]{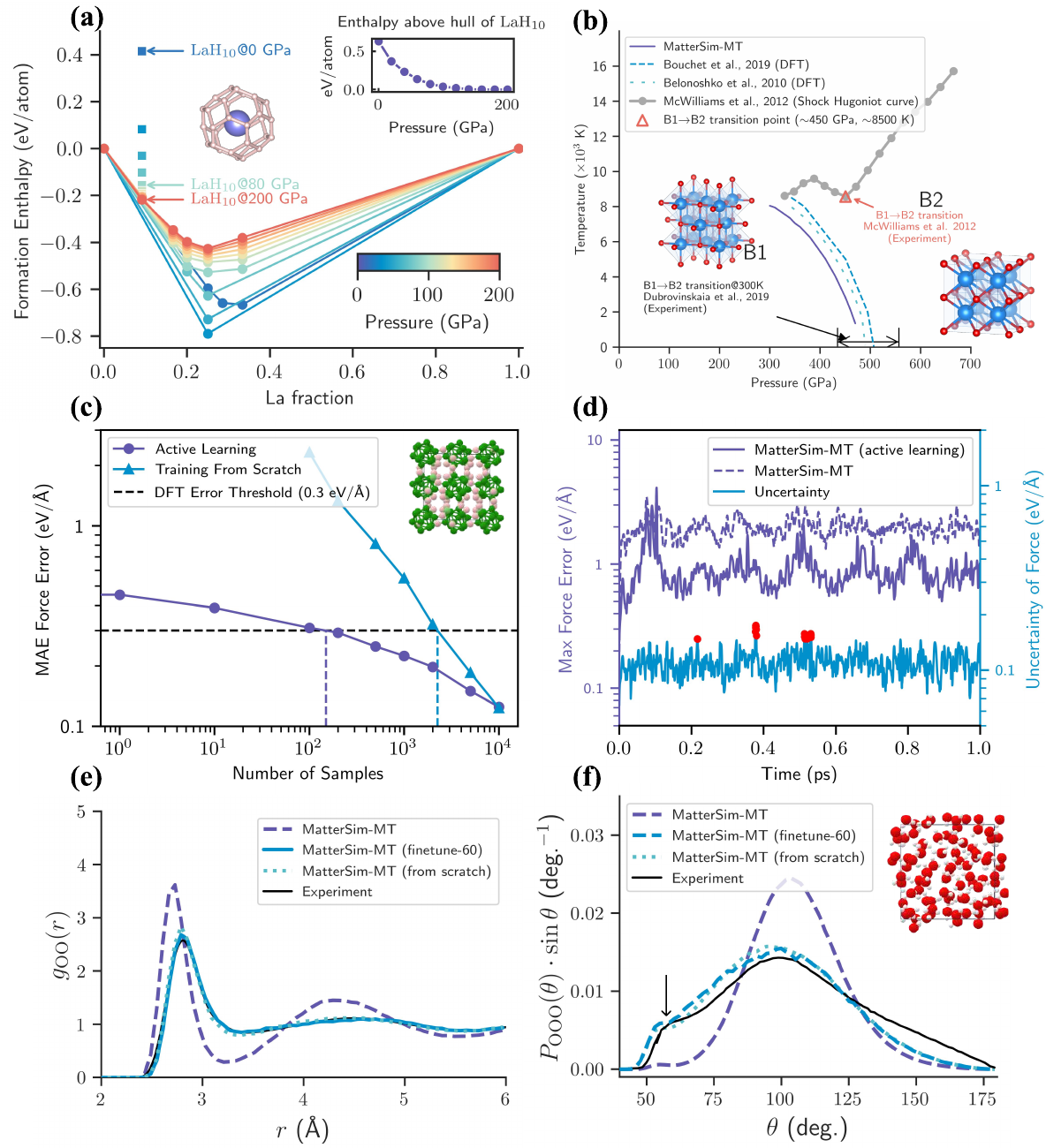}
    \caption{
    \textbf{MatterSim-MT is a foundation model for predicting material structure, dynamics and thermodynamics.}
    (a) Enthalpy above hull analysis of the La-H chemical system at pressures up to \SI{200}{GPa}. Inset shows the stabilization of \ce{LaH10} with increasing pressure.
    (b) Predicted B1-B2 phase boundary of \ce{MgO}, benchmarked against first-principles calculations and experimental data.
    \textbf{(c)} Mean absolute error (MAE) of forces against first-principles results in the simulation trajectory of \ce{Li2B12H12} (crystal structure shown in top right) from an actively learned model and a model trained from scratch, with increasing number of data samples used in active learning.
    \textbf{(d)} Force error and prediction uncertainty along the \textit{ab initio} molecular dynamics (MD) trajectory for \ce{Li2B12H12}. Red points represent structures with above-threshold uncertainty taken for active learning.
    \textbf{(e)} Oxygen-oxygen radial distribution functions (RDF) of bulk water from MD simulations using the pretrained \texttt{MatterSim-MT} model (zero-shot), \texttt{MatterSim-MT} finetuned on 60 water snapshots and \texttt{MatterSim-MT} trained from scratch on 900 water snapshots. Black line indicates the experimental reference.\cite{skinner2014structure,chen2016ab}
    \textbf{(f)} Oxygen-oxygen-oxygen-angular distribution functions, $P_\mathrm{OOO}(\theta)$, from MD simulations using the pretrained \texttt{MatterSim-MT} model, \texttt{MatterSim-MT} finetuned on 60 water snapshots and \texttt{MatterSim-MT} trained from scratch on 900 water snapshots. Black line represents the empirical potential structural refinement (EPSR) of joint X-ray and Neutron scattering measurements of bulk water at \SI{298}{K}.\cite{soper2008quantum}
    }
    \label{fig:placeholder}
    \label{fig:hpht}
    \label{fig:al-and-ft}
\end{figure}

As a materials foundation model pretrained on a large-scale, first-principles PES, \texttt{MatterSim-MT} delivers accurate predictions of energies, forces, and stresses for atomic structures across a wide range of chemical compositions, temperatures, and pressures.
As an illustrative example, \autoref{fig:overview}(d) presents a comparison of predicted stress errors for materials\cite{giannessi2024database} that are predicted to be stable across a broad pressure range according to first-principles calculations. \texttt{MatterSim-MT} reproduces these stress states with a mean error of \SI{8.5}{\giga\pascal}, demonstrating its ability to accurately capture structural responses from ambient to high-pressure conditions.
In this section, we evaluate the model's ability to predict material stability and phase diagrams under finite temperature and pressure conditions. Additional benchmarks on phonon spectra, mechanical properties, and molecular dynamics simulations are provided in \autoref{si-sec:benchmarks}.

First, we evaluate the model's ability to predict material stability. \texttt{MatterSim-MT} demonstrates strong performance in predicting phonon properties, achieving errors of approximately \SI{1}{\tera\hertz} for max phonon frequencies (\autoref{si-tab:phonon-free-energy-results}) and \SI{23}{\kilo\meter/\second} for group velocities (\autoref{fig:overview}(d)).
These capabilities enable accurate estimation of vibrational free energies. In \autoref{fig:overview}(d), we further demonstrate the model’s ability to predict free energy differences between \SIrange{300}{900}{\kelvin} for inorganic crystalline materials listed in \autoref{si-tab:free-energy-benchmark-materials}, achieving a mean absolute error of \SI{18.50}{meV\per atom} compared to first-principles calculations.
Notably, the error drops to \SI{9.62}{meV\per atom} when increasing the number of parameters up to \SI{1.3}{B} (\texttt{MatterSim-MT-1.3B}), further validating the scalability of the model.  This ability enables rapid \textit{ab initio} prediction of material stability under practical conditions.

Next, we evaluate the model's ability to predict phase diagrams. In \autoref{fig:hpht}(a), we focus on the La--H chemical system and study the phase stability of \ce{LaH10}, a well-known high-temperature superconductive hydride previously synthesized under extreme pressure.\cite{drozdov2019superconductivity}
A comparison with several publicly available foundational uMLIPs on the La--H convex hull (\autoref{si-fig:case-study-bench-LaH10}) shows that \texttt{MatterSim-MT} faithfully reproduces stabilization of \ce{LaH10} under pressure at $\sim$\SI{150}{\giga\pascal}, while others either predict premature stabilization or exhibit qualitatively different behavior at elevated pressures. Interestingly, we also observe a destabilization range of a few tens of gigapascals for the \ce{LaH2} phase, which has been verified through first-principles computations.\cite{drozdov2019superconductivity} Traditionally, deriving the pressure-dependent stability range of binary systems via experiments or conventional first-principles computations would take months.
In this example, \texttt{MatterSim-MT} can predict phase stability within minutes.

We also predict the temperature- and pressure-dependent phase diagram of \ce{MgO},\cite{dubrovinskaia2019b1,zhang2023toward} a mineral that shapes the convection of Earth's mantle\cite{tang2010lattice,lay1998core}. Our predictions extend to extreme conditions of \SI{10000}{\kelvin} and \SI{500}{\giga\pascal}, with phase boundaries showing good agreement compared to both experimental measurements\cite{dubrovinskaia2019b1,mcwilliams2012phase} and first-principles density functional theory (DFT) calculations (\autoref{fig:hpht}(b)).\cite{belonoshko2010mgo,bouchet2019ab} A comparison with several publicly available foundational uMLIPs on the MgO B1--B2 phase boundary (\autoref{si-fig:case-study-bench-MgO}) confirms that \texttt{MatterSim-MT} better reproduces the DFT reference, while others fail to capture the temperature dependence and shape of the phase boundary.  These results demonstrate that \texttt{MatterSim-MT} serves as a high-accuracy uMLIP for energy-based simulations, especially at higher temperatures and pressures.

As a materials foundation model, \texttt{MatterSim-MT} can be further extended to new systems via active learning. For complex systems where out-of-distribution configurations may arise during long molecular dynamics simulations,\cite{fu2022forces,xie2023uncertainty,kong2023overcoming} \texttt{MatterSim-MT} employs uncertainty-driven active learning to selectively label and incorporate the most informative configurations. We benchmark this approach on molten phosphorus, molten boron, and lithium dodecahydro-closo-dodecaborate (\ce{Li2B12H12}) (see details in \autoref{si-sec:active-learning}). As shown in \autoref{fig:al-and-ft}(c,d), the actively learned model matches the accuracy of a model trained from scratch while requiring only a fraction of the data---just 6.6\% for \ce{Li2B12H12}---confirming the broad configurational coverage of the pretrained model.

Similarly, the model can be fine-tuned to higher levels of theory beyond the Perdew–Burke–Ernzerhof (PBE) functional\cite{perdew1996generalized} used during pretraining (detailed in \autoref{si-sec:computation-details}). As a demonstration, we fine-tune \texttt{MatterSim-MT} to the rev-PBE0-D3 level of theory\cite{cheng2019ab} for liquid water (see \autoref{si-sec:finetune-water-setups} for details). With only 60 configurations, the fine-tuned model predicts the structural and dynamical properties of water (\autoref{fig:al-and-ft}(e,f)) as accurate as a model trained from scratch on 900 configurations. The fine-tuned model also predicts the self-diffusion coefficient $D=0.193\pm0.011$\SI{}{\square\angstrom\per\ps} within 20\% of the experimental value range \SIrange{0.23}{0.24}{\square\angstrom\per\ps}.\cite{krynicki1978pressure,hardy2001isotope} This corresponds to a 92\% reduction in data generation cost (see \autoref{si-fig:water-finetune}).

\section{MatterSim-MT as a multi-task materials foundation model}

\begin{figure}
    \centering
    \includegraphics[width=\textwidth]{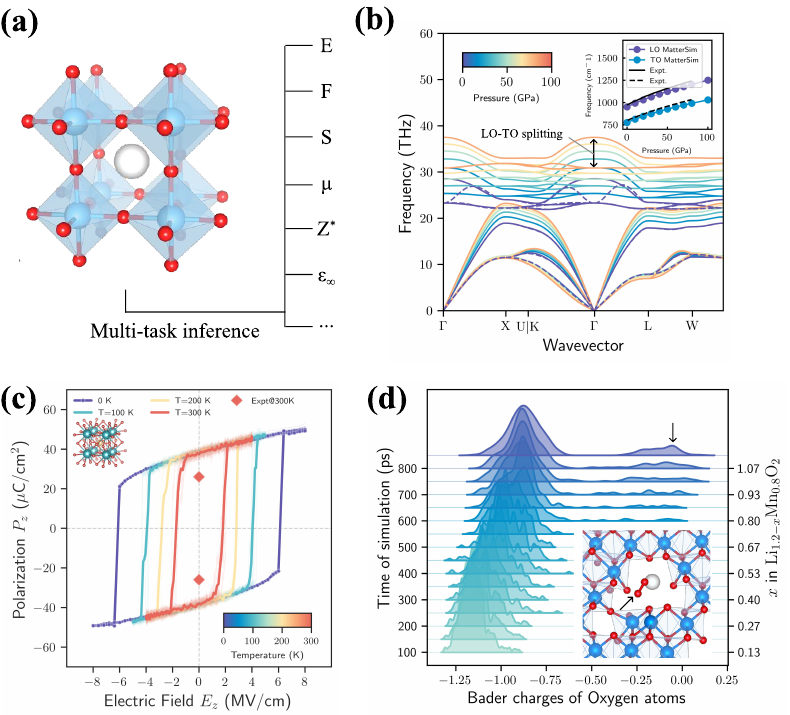}
    \caption{\textbf{MatterSim-MT as a multi-task emulator beyond the potential energy surface.} \textbf{(a)} Illustration of the multi-task inference capabilities of \texttt{MatterSim-MT}, including predictions of energy (E), forces (F), stress (S), magnetic moments ($\mu$), Born effective charges ($Z^*$), and dielectric matrices ($\varepsilon_\infty$) from atomic structures.
    \textbf{(b)} Pressure-dependent phonon spectrum of \ce{SiC} up to \SI{100}{\giga\pascal}, with inset comparing MatterSim's predicted longitudinal optical (LO) and transverse optical (TO) splitting against experimental measurements, and the dashed curve illustrating the phonon dispersion under \SI{0}{\giga\pascal} \textit{without} LO-TO splitting for comparison.
    \textbf{(c)} Predicted hysteresis curve of polarization density as a function of the electrical field along z direction in the ferroelectric tetragonal \ce{BaTiO3} material.
    \textbf{(d)} Evolution of oxygen Bader charge distributions in \ce{Li_{1.2-x}Mn_{0.8}O2} during delithiation, with arrows indicating the formation of an \ce{O2} molecule.}
    \label{fig:multi-task-capabilities}
\end{figure}

Many important physical  phenomena cannot be captured by PES predictions alone. \texttt{MatterSim-MT} natively predicts several other fundamental material properties, including atomic Bader charges, magnetic moments (assuming ferromagnetic ordering), Born effective charges ($Z^*$), and dielectric matrices ($\varepsilon_\infty$). The ability to accurately predict these properties is crucial for applications such as catalysis, energy storage, and planetary sciences.
On held-out test sets, the model achieves mean absolute errors of \SI{0.023}{e} for Bader charges, \SI{0.064}{\mu_B} for magnetic moments, 0.076 $e$ for Born effective charges, and
0.25 for dielectric tensor elements (see \autoref{fig:overview}(d) and \autoref{si-tab:multi-task-results}).
These additional outputs enable simulations of physical phenomena that are inaccessible to models trained on only PES data while spanning a much wider range of tasks than specialized expert models.
We illustrate these multi-task capabilities through three case studies spanning vibrational spectroscopy, ferroelectric switching, and electrochemical redox, each requiring a distinct combination of property predictions.

First, we demonstrate how predictions of Born effective charges and dielectric properties enable the computation of phonon spectra in polar crystals, where long-range Coulomb interaction between ions splits the longitudinal optical (LO) and transverse optical (TO) phonon modes at the Brillouin zone
center.\cite{baroni2001phonons} LO-TO splitting is a hallmark of polar crystals but is absent in standard force-constant approaches. That is because it requires knowledge of the Born effective charge tensor $Z^*$ and the electronic dielectric matrix $\varepsilon_\infty$ to construct the non-analytical correction to the dynamical matrix.\cite{cochran1962dielectric,giannozzi1991ab,baroni2001phonons}
As an illustration, we simulate the LO--TO splitting of phonon modes in 3$c$-\ce{SiC}, a promising material for
high-temperature and high-power electronics, under pressures up to \SI{100}{\giga\pascal}.\cite{neudeck2006silicon}
 As shown in \autoref{fig:multi-task-capabilities}(b) and \autoref{si-tab:sic-bec-and-dielectric}, \texttt{MatterSim-MT} predicts a Born effective charge $Z^*_\mathrm{Si} = 2.71$ $e$
and a dielectric constant $\varepsilon_\infty = 7.31$ at zero pressure, in close agreement with both theoretical ($Z^*_\mathrm{Si} = 2.72$ $e$, $\varepsilon_\infty =
7.02$)\cite{karch1996pressure} and experimental ($Z^*_\mathrm{Si} = 2.70$ $e$, $\varepsilon_\infty = 6.52$)\cite{olego1982pressure-BEC,yu2005fundamentals} values. The resulting LO--TO
splitting of \SI{5.26}{\tera\hertz} deviates by only \SI{0.06}{\tera\hertz} from \textit{ab initio} calculations\cite{karch1996pressure} and \SI{0.03}{\tera\hertz} from experimental
measurements.\cite{olego1982pressure-frequency} The inset of \autoref{fig:multi-task-capabilities}(b) further shows that the predicted pressure dependence of the LO and TO phonon frequencies agrees well with experiments across the full pressure range. Notably, \texttt{MatterSim-MT} accurately predicts Born effective charges and dielectric matrices for pressures up to \SI{100}{\giga\pascal}, even though the training data did not contain any labels at such pressure levels. This result highlights the model's ability to extrapolate beyond its training data.

The predicted Born effective charges also enable the simulation of polarization-driven phenomena under external electric fields. In ferroelectric materials,
the coupling between atomic displacements and the macroscopic polarization gives rise to switchable polarization states. Simulating this behavior requires accurate interatomic forces and Born effective charges that relate atomic displacements to the macroscopic polarization.
For \ce{BaTiO3}, we incorporate the contribution of an external electric field into molecular dynamics simulations and reproduce the polarization-electric field hysteresis curve (\autoref{fig:multi-task-capabilities}(c)). The simulations correctly predict that finite-temperature effects lead to a reduced coercive field, even though the spontaneous polarization of \SI{38}{\mu C\per\centi\meter^2} predicted at \SI{300}{\kelvin} is slightly higher than the experimental value (\SI{26}{\mu C\per \centi\meter^2}), likely due to the well-known underbinding at the PBE level of theory (see Ref.~\citenum{falletta2025unified} and \citenum{martin2025general}). More broadly, the ability to simulate polarization hysteresis without material-specific fitting suggests that this approach could be readily utilized to study complex ferroelectric phenomena such as ferroelectric-paraelectric phase transitions across diverse chemical systems.

Finally, we simultaneously predict atomic charges and magnetic moments to study the electronic degrees of freedom in chemical bonding and redox processes. Here, we examine the behavior of the cathode
material \ce{Li_{1.2-x}Mn_{0.8}O2} under electrochemical charging conditions. These lithium-rich transition-metal oxides are promising next-generation batteries due to their high energy density, but suffer from irreversible capacity
loss associated with the anionic oxygen redox mechanism.\cite{mccoll2024phase} We perform molecular dynamics simulations of progressive lithium extraction at \SI{1000}{\kelvin}, removing one lithium atom every \SI{5}{\pico\second} until complete delithiation,
while monitoring ion dynamics, magnetic moments, and atomic Bader charges predicted by \texttt{MatterSim-MT} (see \autoref{si-fig:case-study-li-battery} for details).

Our analysis shows a clear transition from cationic to anionic redox during lithium extraction. At low degrees of delithiation, Mn serves as the primary redox center, with magnetic moment changes of approximately \SIrange{3}{4}{\mu_B}. Upon reaching $x \approx 0.5$ in \ce{Li_{1.2-x}Mn_{0.8}O2}, Mn ions begin migrating from octahedral sites in the transition-metal layers to tetrahedral sites in the lithium layers, accompanied by distortions of the oxygen sublattice. Further delithiation to $x \approx 0.9$ triggers the
formation of oxygen dimers, typically following Mn migration events. The evolution of these unstable dimers leads to the formation of molecular \ce{O2} within the lattice (\autoref{fig:multi-task-capabilities}(d), inset), a hallmark
of irreversible anionic redox. This is corroborated by the Bader charge analysis in \autoref{fig:multi-task-capabilities}(d), where the \ce{O2} molecules
exhibit near-zero charges, confirming their molecular rather than ionic character. Continued delithiation destabilizes the lattice further, resulting in increased molecular oxygen
formation and an eventual transition toward nanofluidic oxygen states.\cite{mccoll2024phase} Notably, this comprehensive picture of the cationic-to-anionic redox transition and lattice degradation naturally emerges from the multi-task predictions, without any task-specific training on battery materials.

\section{Discussion}
With the advent of efficient generative models for materials exploration, accurate property characterization is becoming a major bottleneck in computational materials design.
Atomistic foundation models offer a promising route to accelerate property characterization, but previous models have mostly focused on modeling the PES.
In this work, we introduce \texttt{MatterSim‑MT}, a multi‑task foundation model for \textit{in silico} materials property characterization beyond the PES. The model enables atomistic simulations for materials composed of arbitrary elemental combinations, for temperatures up to \SI{5000}{\kelvin} and pressures up to \SI{1000}{\giga\pascal}, providing accurate predictions for energetics, vibrational properties, and thermal transport. \texttt{MatterSim-MT} also natively predicts atomic Bader charges, magnetic moments, Born effective charges, and dielectric matrices, enabling simulations under external conditions such as electric fields using a single model. Although we demonstrate multi-task prediction for four properties beyond the PES, the underlying framework is general: Additional properties computed from first principles calculations can be incorporated through the same strategy.
As the availability of high-quality synthetic data continues to grow, this multi-task paradigm allows us to greatly expand the range of applications that atomistic foundation models can address.

Despite these advances, several limitations remain. Scaling the model size increases computational cost, making system-level solutions such as model compression and efficient system engineering increasingly important.\cite{choudhary2020comprehensive,wang2023bitnet,ding2023longnet}
More fundamentally, the  physical reliability of predicted properties is limited by the underlying level of theory. For example, Born effective charges and dielectric tensors computed via Berry-phase approaches\cite{king1993theory,resta1994macroscopic} become ill-defined in narrow-gap or metallic systems,\cite{souza2002first} and a multi-task model may inadvertently learn these nonphysical divergences. Relatedly, some properties may not be well-defined for some systems. For example, we predict magnetic density assuming ferromagnetic ordering, but the material may not be ferromagnetic in practice. These examples highlight the need for careful data curation, awareness of method failure modes, and task-dependent validation as the number of supported properties grows.

\backmatter

\section*{Acknowledgements}
We thank Chris Bishop, Tao Qin, Tie-Yan Liu, Haiguang Liu, Bin Shao, Jia Zhang, and Karin Strauss for discussions that helped shape this work; Prof.\ Davide Donadio for comments on applications of MatterSim; and Bichlien Nguyen, Yu Xie, and Jonas K\"ohler for manuscript feedback. We also thank Ryota Tomioka for DFT computation pipelines, and Maik Riechert, Hannes Schulz, Stefano Battaglia, and Thijs Vogels for computational infrastructure support. We appreciate Shoko Ueda, Peggy Dai, and Roberto Sordillo for project management, and Lina Lu, Yang Ou, and Jingyun Bai for graphic design. We are grateful to the Microsoft Research AI for Science team for ongoing discussions. Z.L.\ thanks Pascal Salzbrenner and Prof.\ Chris Pickard for discussions on high-pressure crystal structures.

\section*{Contributions}
\textbf{Conceptualization:} H.Y., J.L., H.H., Z.L.;
\textbf{Data curation:} H.Y., J.L., G.L., Q.W., Y.Zhu, Z.C., S.C.;
\textbf{Methodology:} H.Y., X.L., C.H., Y.Zhou, Y.S., C.Liu, J.T., J.L., H.H., Z.L.;
\textbf{Validation:} H.Y., J.L., G.L., Q.W., Y.Zhu, Z.C., S.C., F.T., C.Z., M.H., R.P., A.F., D.Z., T.X., J.S., L.S., Y.Chen, L.K., H.H., Z.L.;
\textbf{Software:} Y.B., D.G.;
\textbf{Supervision:} H.Y., F.N., H.H., Z.L.;
\textbf{Resources:} Z.L.;
\textbf{Writing -- review \& editing:} All authors.

\section*{Inclusion and ethics declarations}
\textbf{Competing interests}:
The authors declare no competing interests.

\section*{Data availability}
The benchmark datasets will be made publicly available following editorial request upon publication.

\section*{Code availability}
We will open-source the code and weights following editorial request upon publication.

\bibliography{ms}

\end{document}


\title{Supplementary Information}

\affil*[1]{\orgname{Microsoft Research AI for Science}}

\maketitle

\tableofcontents

\clearpage

\section{Model architecture and training details}\label{sec:model-arch}
\subsection{Materials Graphs}
\begin{figure}
    \centering
    \includegraphics[width=1\linewidth]{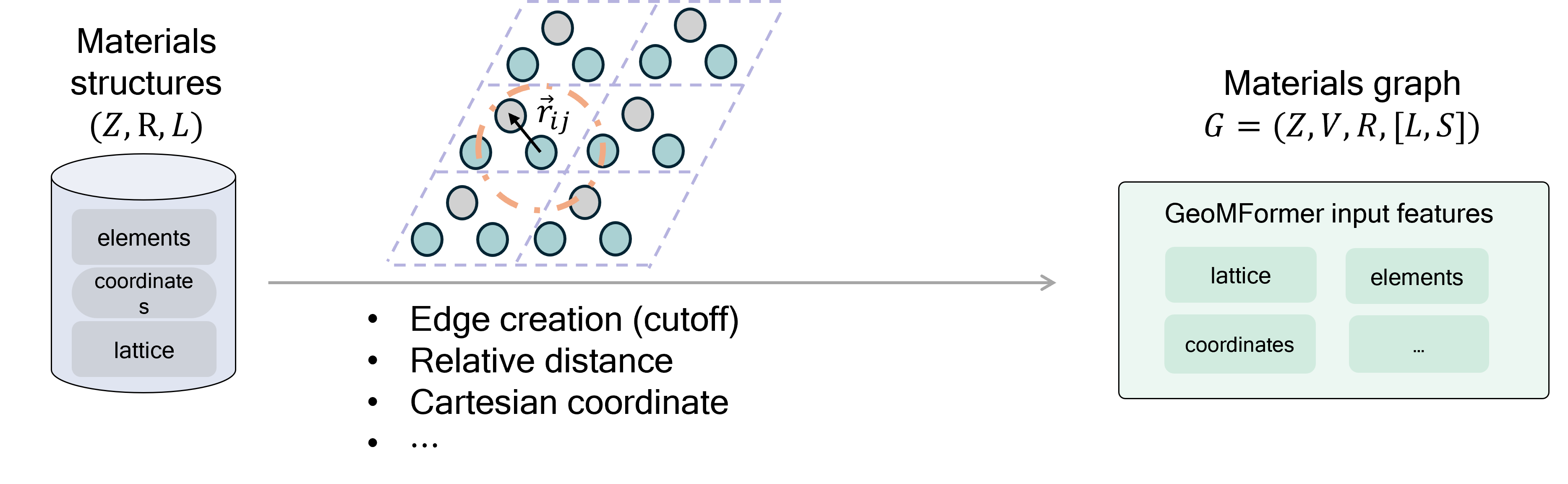}
    \caption{MatterSim leverages materials graphs built upon point clouds to represent atomic interactions and geometric features in Euclidean space.  }
    \label{fig:mattersim-materials-graph}
\end{figure}

MatterSim operates on materials graphs constructed from three-dimensional point clouds under periodic boundary conditions. We define a materials graph $\mathcal{G} = (\boldsymbol{Z},\boldsymbol{V},\boldsymbol{R},[\boldsymbol{L}, \boldsymbol{S}])$ (see \autoref{fig:mattersim-materials-graph}), where $\boldsymbol{Z}$ denotes atomic numbers and additional per-atom features, $\boldsymbol{R}$ contains atomic positions $\boldsymbol{r}_i \in \mathbb{R}^{3}$, $\boldsymbol{V}$ encodes relative displacement vectors between atom pairs, $\boldsymbol{L}$ is the $3\times 3$ lattice matrix, and $\boldsymbol{S}$ represents global scalar state variables such as temperature and pressure. Nodes correspond to atoms and edges are formed between all pairs within a radial cutoff distance $r_c$. Fractional coordinates are converted to Cartesian before graph construction. By design, MatterSim preserves roto-translational invariance for scalar properties (e.g., total energy) and equivariance for vectorial properties (e.g., forces).

\subsection{Model Architecture}
\texttt{MatterSim-MT} adopts a transformer-based architecture inspired by GeoMFormer.\cite{chen2023geomformer} A separate M3GNet-based \texttt{MatterSim-v1}\cite{yang2024mattersim} model is also utilized to drive the materials explorer during structure sampling.

The model is built on the Transformer architecture\cite{vaswani2017attention} and adapted to leverage SE(3)-equivariant vectors for both molecular and materials systems. It employs two parallel streams that maintain and learn invariant and equivariant representations, respectively. Following GeoMFormer,\cite{chen2023geomformer} a cross-attention module bridges the two streams, enabling information exchange and improving geometric modeling.

To handle the periodic boundary conditions (PBC) of crystal structures, we incorporate the multi-graph construction from Ref.~\citenum{xie2018crystal}, representing atoms in the unit cell via periodic graphs that include image atoms from neighboring cells up to a predefined cutoff distance. A smooth cutoff function based on interatomic distances gradually attenuates long-range interactions, ensuring translational invariance.

\begin{figure}
    \centering
    \includegraphics[width=0.75\linewidth]{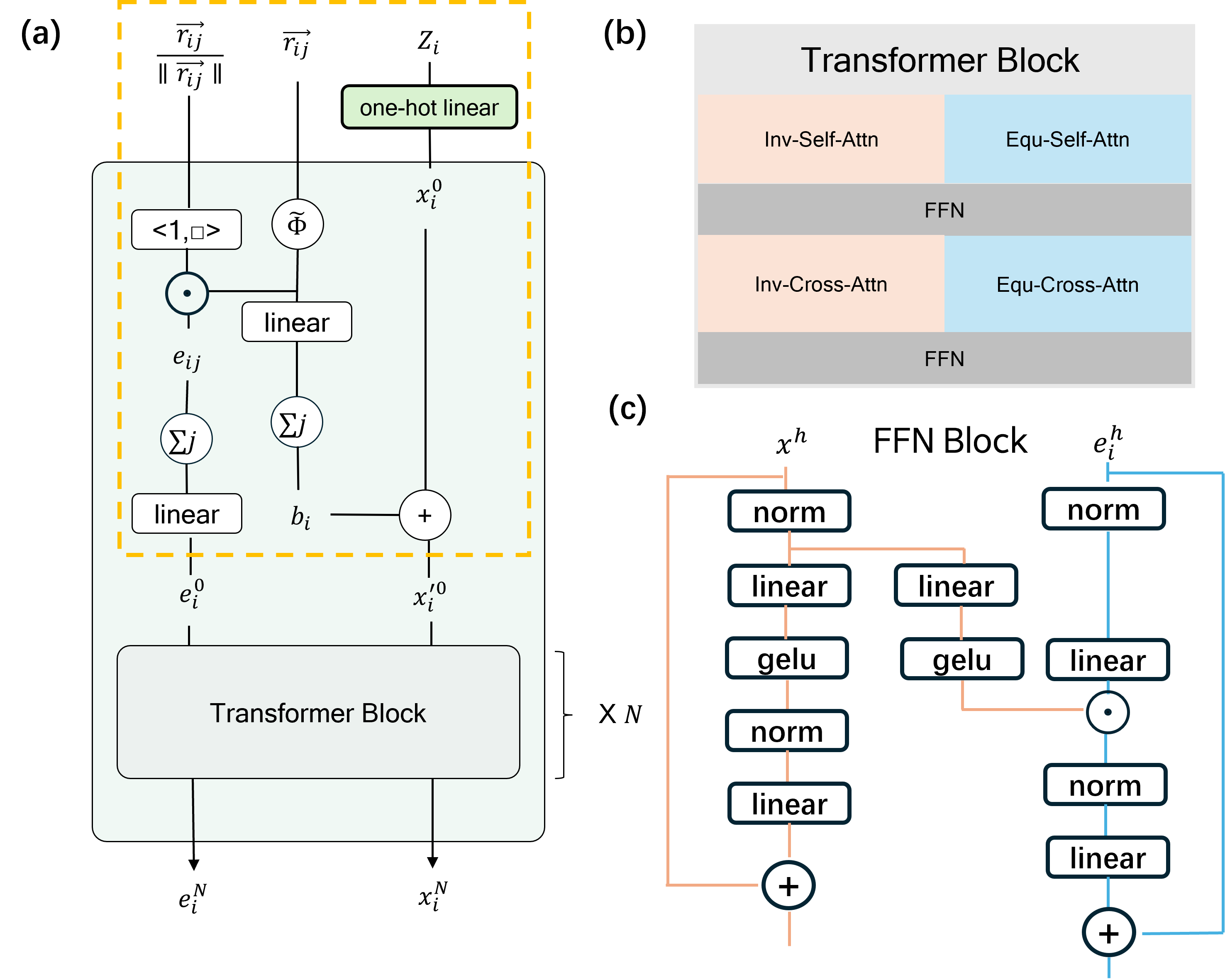}
    \caption{(a) The overall architecture of the model, which extracts atomic embeddings using the embedding block and Transformer block. The yellow line highlights the embedding block. (b) The Transformer block comprises Self-Inv-Attn, Self-Equ-Attn, Cross-Inv-Attn, Cross-Equ-Attn, and the FFN module. (c) Implementation of the FFN module.}
    \label{fig:model-overview}
\end{figure}

\textbf{Embedding Block.} The model architecture is shown in \autoref{fig:model-overview}(a), with the Embedding Block highlighted in yellow. It consists of three components: atom embedding, spatial relationship embedding, and equivariant feature embedding. Atom embedding maps atomic species $Z_i$ and positions $\boldsymbol{r}_i$ into feature vectors $\boldsymbol{x}_i$, initialized via a one-hot linear layer.

The spatial relationship embedding captures interatomic relationships through centrality encoding:

\begin{equation}
\boldsymbol{b}_i = \sum_{j\in\mathcal{N}(i)}\mathrm{Linear}\left(m_{ij} \cdot \tilde{\Phi}(\norm{\boldsymbol{r}_{ij}})\right)
\end{equation}

where $\mathcal{N}(i)$ denotes the neighbors of atom $i$ within cutoff radius $r_c$, $\boldsymbol{r}_{ij} = \boldsymbol{r}_i - \boldsymbol{r}_j$ is the relative displacement, and $\tilde{\Phi}(\cdot)$ is a Gaussian basis kernel. The smooth cutoff mask $m_{ij}$ is defined as:

\begin{equation}
m_{ij} = 1 - 6 \left(\frac{\norm{\boldsymbol{r}_{ij}}}{r_c}\right)^5 + 15 \left(\frac{\norm{\boldsymbol{r}_{ij}}}{r_c}\right)^4 - 10 \left(\frac{\norm{\boldsymbol{r}_{ij}}}{r_c}\right)^3
\end{equation}

The centrality encoding updates the initial embedding as $\boldsymbol{x}_i^{\prime 0} = \boldsymbol{x}_i^0 + \boldsymbol{b}_i$, which is then passed to the attention modules.

The equivariant feature embedding captures relative positional information for the equivariant stream. It is initialized as:

\begin{equation}
\boldsymbol{e}_i^0 = \sum_j m_{ij} \tilde{\Phi}(\norm{\boldsymbol{r}_{ij}}) \cdot \text{concat}\left(1,\frac{\boldsymbol{r}_{ij}}{\norm{\boldsymbol{r}_{ij}}}\right)
\end{equation}

where $\text{concat}\left(1,\frac{\boldsymbol{r}_{ij}}{\norm{\boldsymbol{r}_{ij}}}\right)$ concatenates a scalar 1 with the unit displacement vector.

\textbf{Transformer Block.} Each Transformer block contains four attention modules---Inv-Self-Attn, Equ-Self-Attn, Inv-Cross-Attn, and Equ-Cross-Attn---as shown in \autoref{fig:model-overview}(b). Self-attention modules derive key--value pairs from their own stream, while cross-attention modules exchange information between the two streams.

\begin{figure}
    \centering
    \includegraphics[width=\linewidth]{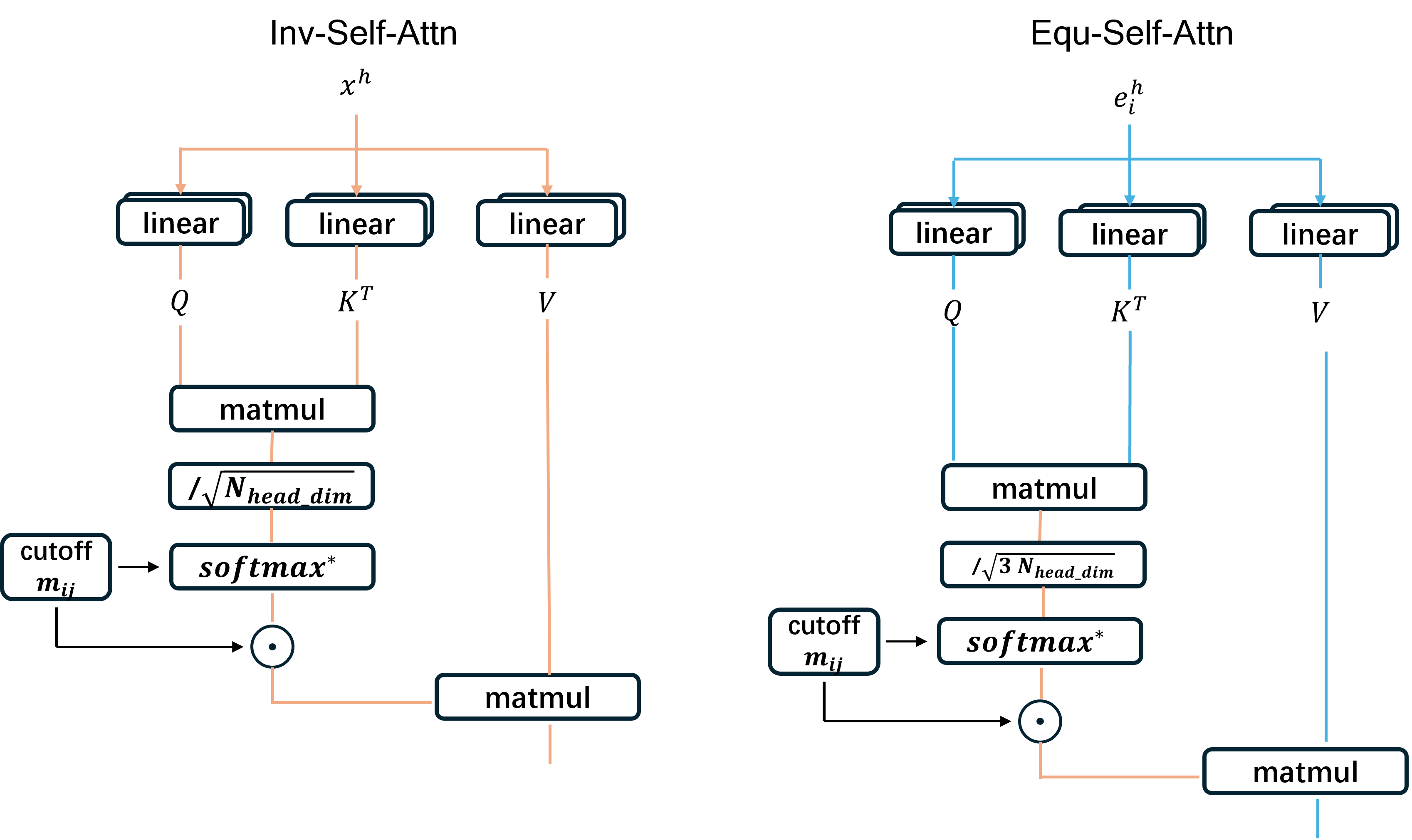}
    \caption{Implementation of the Inv-Attention module.}
    \label{fig:Inv-Attn}
\end{figure}

\begin{figure}
    \centering
    \includegraphics[width=\linewidth]{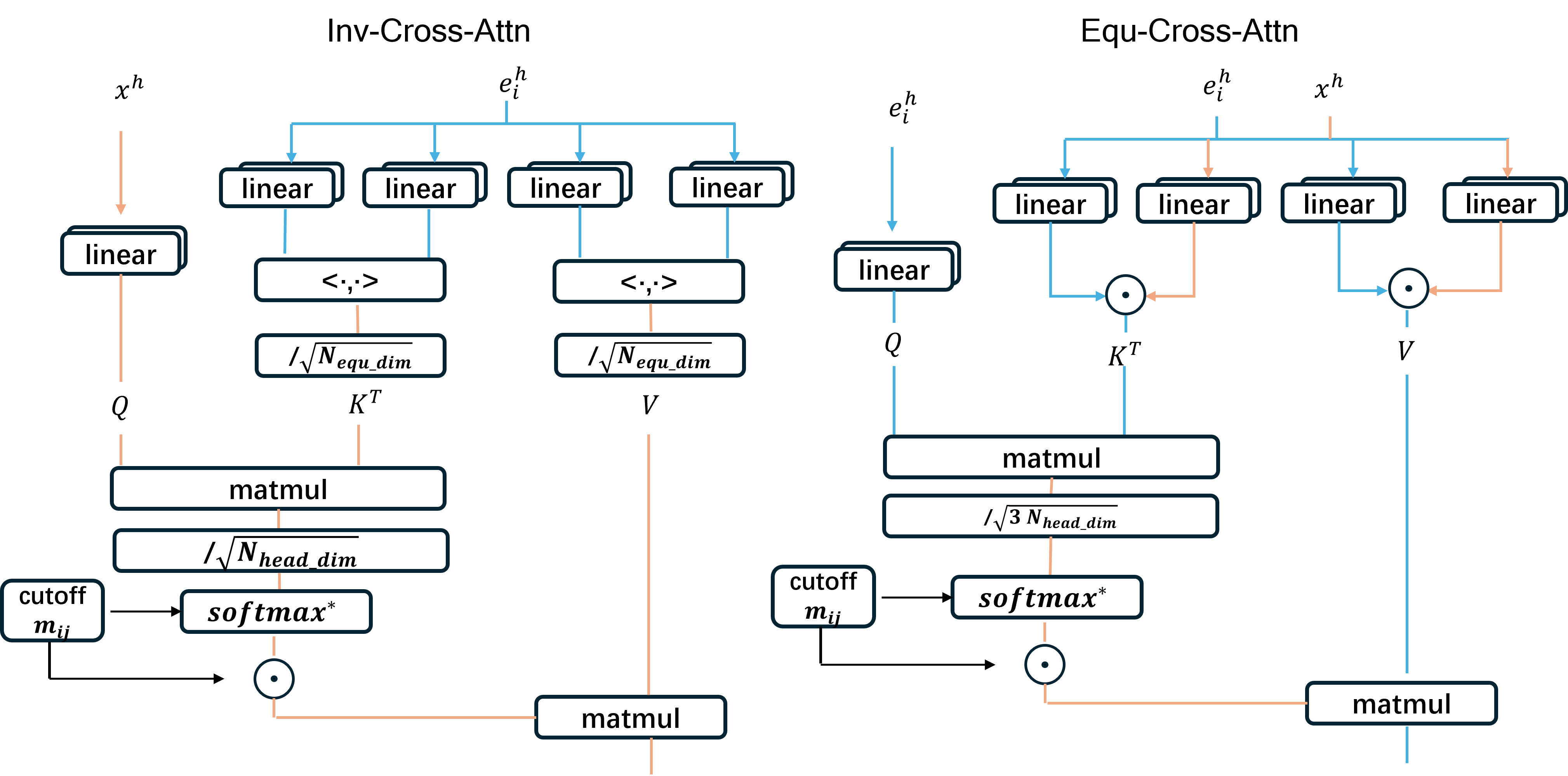}
    \caption{Implementation of the Cross-Attention module.}
    \label{fig:Cross-Attn}
\end{figure}

The FFN module is shown in \autoref{fig:model-overview}(c). Invariant features $x_i$ are updated via a standard feed-forward network with residual connections. Because non-linear activations would break equivariance, equivariant features $e_i$ are instead gated by invariant features, following the GeoMFormer design.

For the self-attention modules at the $h$-th layer (see \autoref{fig:Inv-Attn}), we illustrate with Inv-Self-Attn. The query $\boldsymbol{Q}$, key $\boldsymbol{K}$, and value $\boldsymbol{V}$ matrices are obtained via linear projections of $\boldsymbol{x}^h_i$:

\begin{equation}
\boldsymbol{a}_{ij} = \left(\frac{\boldsymbol{Q}\boldsymbol{K}^T}{\sqrt{d}}\right)_{i,j}
\end{equation}

\begin{equation}
\mathrm{Softmax^*} (\boldsymbol{a}_{ij}) = \frac{e_{ij}^{a_{ij}} \cdot m_{ij}}{\sum_{k\in\mathcal{N}(i)} e_{ik}^{a_{ik}} \cdot m_{ik}}
\end{equation}

\begin{equation}
\boldsymbol{x}_i^{h+1} = \sum_{j\in\mathcal{N}(i)} \mathrm{Softmax^*}\left(\boldsymbol{a}_{ij}\right) \cdot m_{ij} \cdot \boldsymbol{V}_j
\end{equation}

Here, $d$ is the hidden dimension. Note that $\boldsymbol{K}$ and $\boldsymbol{V}$ for periodic image atoms are copied from the original atoms to ensure consistent representations between an atom and its images.

The cross-attention module (see \autoref{fig:Cross-Attn}) enables interaction between the two streams. Its $\boldsymbol{Q}$, $\boldsymbol{K}$, $\boldsymbol{V}$ computation follows the GeoMFormer design, and the attention mechanism is identical to that of Inv-Self-Attn.

\textbf{Task Head.} The Transformer block produces both invariant and equivariant per-atom features, which are used by task-specific heads to predict target properties.

Each task head receives the atomic-level invariant features $x^N_i$ and equivariant features $e^N_i$ from the final Transformer block. These are processed through an Inv-Cross-Attn layer to produce updated scalar features $x^{N+1}_i$. For scalar predictions (energy, Bader charge, magnetic moments):

\begin{equation}
\boldsymbol{x}_i^{N+1} = \text{Inv-Cross-Attn}(x_i^N, e_i^N)
\end{equation}

\begin{equation}
\boldsymbol{p}_i = W_2 f_{LN}(\sigma(W_1 x^{N+1}_i + b_1)) + b_2
\end{equation}

Here, $f_{LN}$ is layer normalization and $\sigma$ is the GELU activation. For energy prediction, mean pooling is applied over atoms; for Bader charge and magnetic moment predictions, the per-atom values $p_i$ are used directly.

For Born effective charges and dielectric properties, we follow the ETGNN design (Ref.~\citenum{zhong2022edge}). The dielectric tensor is:

\begin{equation}
\varepsilon = \frac{1}{N}\sum_{i=1}^N \sum_{j\in\mathcal{N}(i)}(\boldsymbol{p}_i \boldsymbol{p}_j) \frac{\vec{r_{ji}}}{{\norm{r_{ji}}}} \otimes \frac{\vec{r_{ji}}}{{\norm{r_{ji}}}}
\end{equation}

where $\boldsymbol{p}_j$ is computed using separate linear layer parameters from $\boldsymbol{p}_i$.

The Born effective charge is decomposed into symmetric and non-symmetric contributions:

\begin{equation}
\boldsymbol{Z}_i = \boldsymbol{Z}_i^{sym} + \boldsymbol{Z}_i^{non-sym}
\end{equation}

\begin{equation}
\boldsymbol{Z}_i^{sym} = \sum_{j\in\mathcal{N}(i)}(\boldsymbol{p}_i \boldsymbol{p}_j)\frac{\vec{r_{ji}}}{{\norm{r_{ji}}}}\otimes\frac{\vec{r_{ji}}}{{\norm{r_{ji}}}}
\end{equation}

\begin{equation}
\boldsymbol{Z}_i^{non-sym} = \sum_{k,j\in\mathcal{N}(i), k \neq j}(\boldsymbol{p}_i \boldsymbol{p}_j \boldsymbol{p}_j)\frac{\vec{r_{ki}}}{{\norm{r_{ki}}}}\otimes\frac{\vec{r_{ij}}}{{\norm{r_{ij}}}}
\end{equation}

\subsection{Training details}
The loss function for the GeoMFormer model is:

\[
L = l(e, e_\mathrm{DFT}) + \omega_f l(\boldsymbol{f}, \boldsymbol{f}_\mathrm{DFT}) + \omega_\sigma l(\boldsymbol{\sigma}, \boldsymbol{\sigma}_\mathrm{DFT}) + \sum_{i=1}^{4} \omega_{t_i} l(\boldsymbol{p}_{t_i}, \boldsymbol{p}_{t_i,\mathrm{DFT}})
\]

where $l(\cdot, \cdot)$ is the mean absolute error (MAE), $e$ is the energy per atom, $\boldsymbol{f}$ is the per-atom force vector, $\boldsymbol{\sigma}$ is the stress tensor, and $\omega_f$, $\omega_\sigma$ are their respective loss weights. The terms $p_{t_i}$ correspond to four auxiliary targets: magnetic moments, Bader charges, Born effective charges, and dielectric matrices ($i \in \{1, \ldots, 4\}$).

\autoref{tab:Mattersim-Hyperparameters} lists the model hyperparameters for the \SI{1}{M}, \SI{10}{M}, \SI{220}{M}, and \SI{1.3}{B} parameter variants. The FFN embedding size refers to the dimensionality within the FFN block. To limit memory consumption during training, pbc\_expanded\_num\_cell\_per\\\_direction and pbc\_expanded\_token\_cutoff control the number of unit cell expansions and the maximum number of atoms after expansion, respectively.

\autoref{tab:Pretrain-parameters} presents the pretraining hyperparameters. After warm-up, the learning rate decays linearly to zero.

\begin{table}[h]
\centering
\caption{Hyperparameters for training MatterSim with GeoMFormer architecture and parameters of \SI{1}{M}, \SI{10}{M}, \SI{220}{M} and \SI{1.3}{B}.}
\begin{tabular}{lcccc}
\toprule
\textbf{Hyperparameters} & \textbf{\SI{1}{M}} & \textbf{\SI{10}{M}} & \textbf{\SI{220}{M}} & \textbf{\SI{1.3}{B}} \\
\midrule
Number of Gaussian Basis                  & 64 & 128 & 128   & 128   \\
Number of Transformer Blocks              & 2   & 3   & 6   & 9   \\
Number of Attention Blocks                     & 8   & 12   & 24   & 36   \\

\midrule
Atom Embedding Size                 & 128 & 320 & 768   & 1536   \\
Hidden Embedding Size               & 128 & 320 & 768   & 1536   \\
FFN Embedding Size                  & 128 & 320 & 3078   & 6144   \\
\midrule
Number of Attention Heads                   & 8 & 16 & 32   & 32   \\
\midrule
Cutoff Radius (\AA)                                  & 5.0 & 5.0 & 5.0 & 5.0 \\
pbc\_expanded\_num\_cell\_per\_direction       & 5 & 5 & 5 & 5 \\
pbc\_expanded\_token\_cutoff       & 256 & 256 & 256 & 256 \\
\bottomrule
\end{tabular}
\label{tab:Mattersim-Hyperparameters}
\end{table}

\begin{table}[h]
\centering
\caption{Pretraining Hyperparameters}
\begin{tabular}{lc}
\toprule
\textbf{Pretraining parameters}  \\
\midrule
optimizer               & AdamW\cite{loshchilov2017decoupled} \\
max learning rate                     & 5e-5    \\
weight decay                            & 0.0 \\
$\beta _1$                              & 0.9 \\
$\beta _2$                              & 0.999 \\
$\epsilon$                              & 1e-8 \\
\midrule
warmup epochs                                 & 1 \\
training epochs                              & 23 \\
early stop epochs                           & 3 \\
\midrule
$\omega_f$                            & 10.0   \\
$\omega_\sigma$                         & 0.07    \\
$\omega_{t_1}$                          & 1.0    \\
$\omega_{t_2}$                       & 1.0    \\
$\omega_{t_3}$                          & 1.0    \\
$\omega_{t_4}$                           & 1.0    \\
\midrule
batch size                             & 1024    \\
EMA                                      & 0.99    \\
\bottomrule
\end{tabular}
\label{tab:Pretrain-parameters}
\end{table}

\begin{figure}
    \centering
    \includegraphics[width=0.5\linewidth]{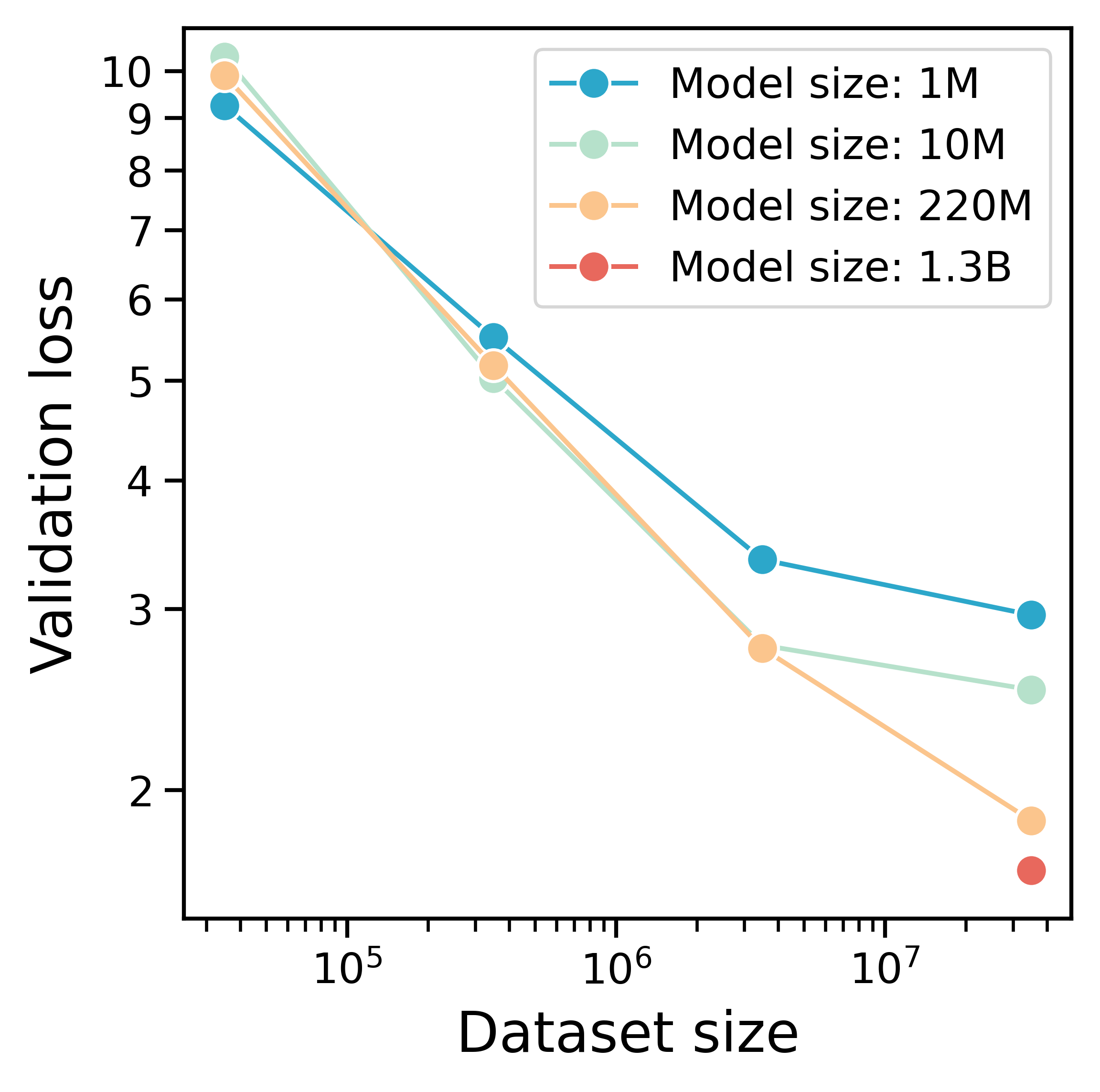}
    \caption{Caption}
    \label{fig:model-scaling}
\end{figure}

\autoref{fig:model-scaling} shows the validation loss as a function of training data size for models of varying capacity. Models with fewer than 10M parameters saturate quickly, whereas larger models continue to improve as the dataset scales up to 35M samples. We further demonstrate scaling to 1.3B parameters; however, given the trade-off between computational cost and accuracy, all predictions and simulations in the main text use the 10M-parameter model.

\section{Materials explorer}\label{sec:materials-explorer}
\subsection{Data exploration}
MatterSim's predictive capabilities are underpinned by a two-part materials structure explorer developed in Ref.~\citenum{yang2024mattersim} that enhances the training datasets through both equilibrium and off-equilibrium structural data.

The \textbf{ground-state explorer} focuses on materials at or near atomistic equilibrium positions. It primarily utilizes an uncertainty-based method with ensemble models to selectively incorporate data from both public repositories and internally generated datasets into the database. This approach ensures the inclusion of the most informative structures for model training, enhancing the accuracy of property predictions at equilibrium. The \textbf{off-equilibrium explorer} targets materials with off-equilibrium atomistic positions. It conducts molecular dynamics (MD) simulations under a wide range of pressures, including 0, 200, 500, 800, and \SI{1000}{\giga\pascal}. Under each pressure, we simulate each material at 300, 1000, 2000, and \SI{5000}{\kelvin}. For each temperature and pressure, the NPT simulation runs for \SI{10}{\pico\second}.

These simulations are crucial for sampling a wide range of atomic configurations, allowing MatterSim to learn and predict material properties under high-pressure and high-temperature scenarios that deviate significantly from equilibrium states.

\subsection{Uncertainty evaluation}\label{sec:materials-explorer-uncertainty}
Uncertainty quantification is essential in predictive modeling for materials properties and simulations, particularly those involving MLFFs. Accurate uncertainty assessment provides insights into the reliability of model outputs and informs decision-making processes. In materials science, where innovation potential is vast but the cost of errors is high, it is essential to estimate the uncertainty of model predictions to ensure reliable outcomes. Current uncertainty quantification methods often involve statistical techniques, such as Bayesian approaches, bootstrapping, and ensemble methods,\cite{tan2024enhanced,krajewski2022extensible,thomas2023calibration,thaler2023scalable} to estimate confidence intervals or prediction errors. These approaches help understand the limits of model predictions and highlight areas needing further refinement or training.

In the case of MatterSim, uncertainty quantification is addressed through an ensemble approach. By training a set of five distinct models with different random initializations, the ensemble of models gives an estimation of the uncertainty on both the energies and forces. The forces offer insight into the dynamical behavior of atoms and can be particularly revealing in scenarios where the predictions for only a small fraction of the atoms within the simulation cell are deemed unreliable. Conversely, energy predictions are often more informative in cases of crystals with a small number of atoms in the cell. By integrating both energies and forces into the uncertainty analysis, MatterSim ensures a robust and reliable assessment of uncertainties, enhancing the confidence in its predictive capabilities for material properties and simulations.

\subsection{First-principles computation details}\label{sec:computation-details}
The DFT parameters employed in this work are generated with the \texttt{MPRelaxSet} class defined in the \texttt{pymatgen} library~\cite{ong2013python} and the calculations are conducted with Vienna Ab-initio Simulation Package (VASP) \cite{kresse1996efficiency,kresse1996efficient} using the projector augmented wave (PAW) method\cite{blochl1994projector} and Perdew-Burke-Ernzerhof (PBE)~\cite{perdew1996generalized} exchange-correlation functional.
To account for on-site electronic repulsion, Hubbard U parameters were applied to the following elements in oxides and fluorides:
\begin{itemize}
\item Co: \SI{3.32}{eV}
\item Cr: \SI{3.7}{eV}
\item Fe: \SI{5.3}{eV}
\item Mn: \SI{3.9}{eV}
\item Mo: \SI{4.38}{eV}
\item Ni: \SI{6.2}{eV}
\item V: \SI{3.25}{eV}
\item W: \SI{6.2}{eV}
\end{itemize}
The cutoff of the plane-wave basis set is \SI{520}{eV} and the convergence threshold for total energy is \SI{5e-5}{eV\per atom}.
For each material, the total energy, forces on each atom and the stress are computed, stored and used for training.
We encountered convergence difficulties with elements such as \ce{Gd} and \ce{Eu}, particularly in off-equilibrium structures where self-consistent cycles failed to converge, or energies varied significantly for two structures with nearly identical atomic positions. Such calculations are consequently excluded from the study.

\section{Training Data}\label{sec:training-data}

\subsection{Off-equilibrium materials data}
In this section, we analyze the distribution of the dataset across the temperature and pressure space, and compare it with public datasets, including \texttt{MPF2021},\cite{chen2022universal} \texttt{MPtrj},\cite{deng2023chgnet}, \texttt{Alexandria}\cite{schmidt2023machine} and \texttt{OMat24}.\cite{barroso2024open}

\begin{figure}
    \centering
    \begin{subfigure}{0.45\textwidth}
    \includegraphics[width=\textwidth]{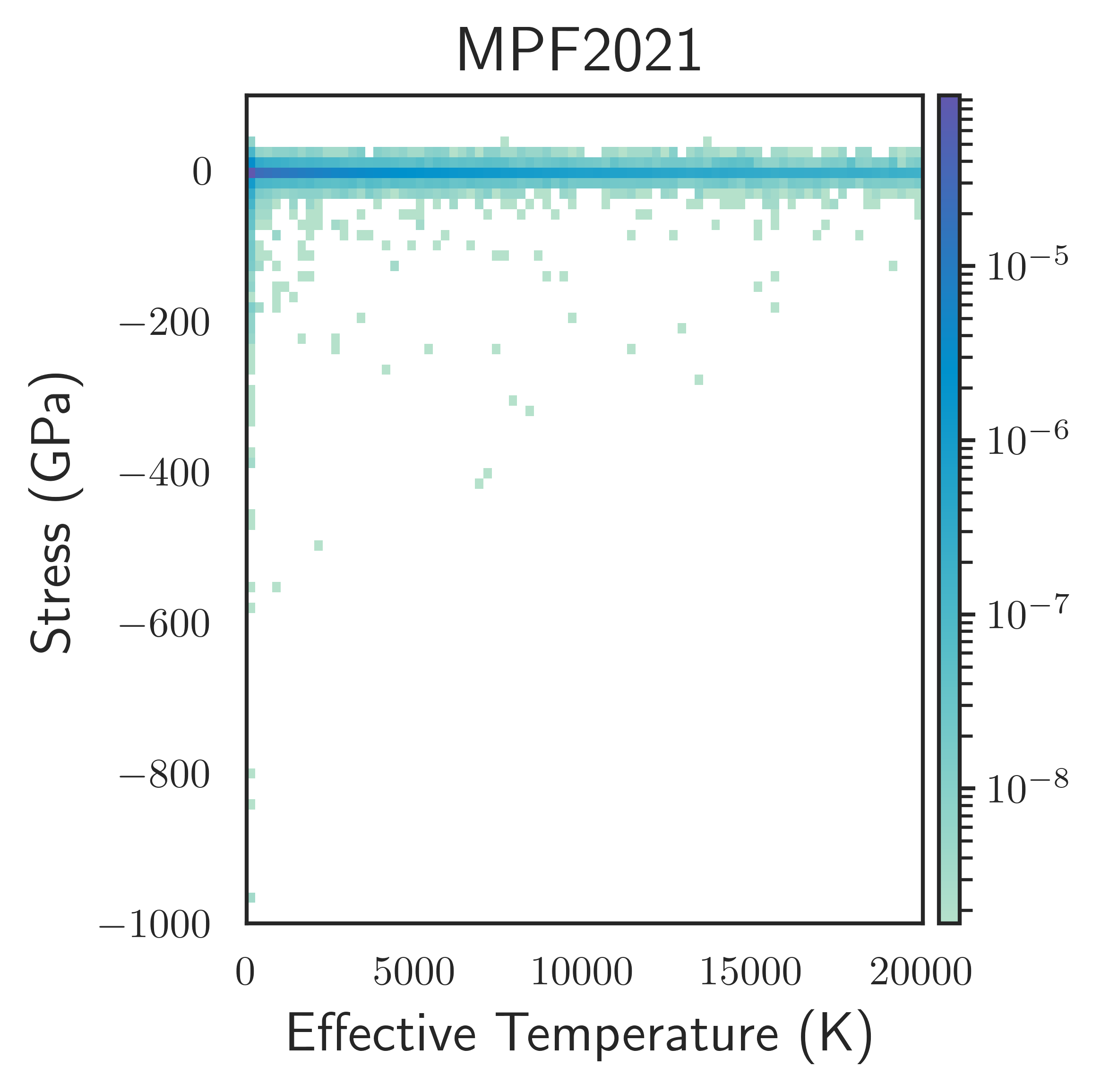}
    \caption{}
    \end{subfigure}
    \begin{subfigure}{0.45\textwidth}
    \includegraphics[width=\textwidth]{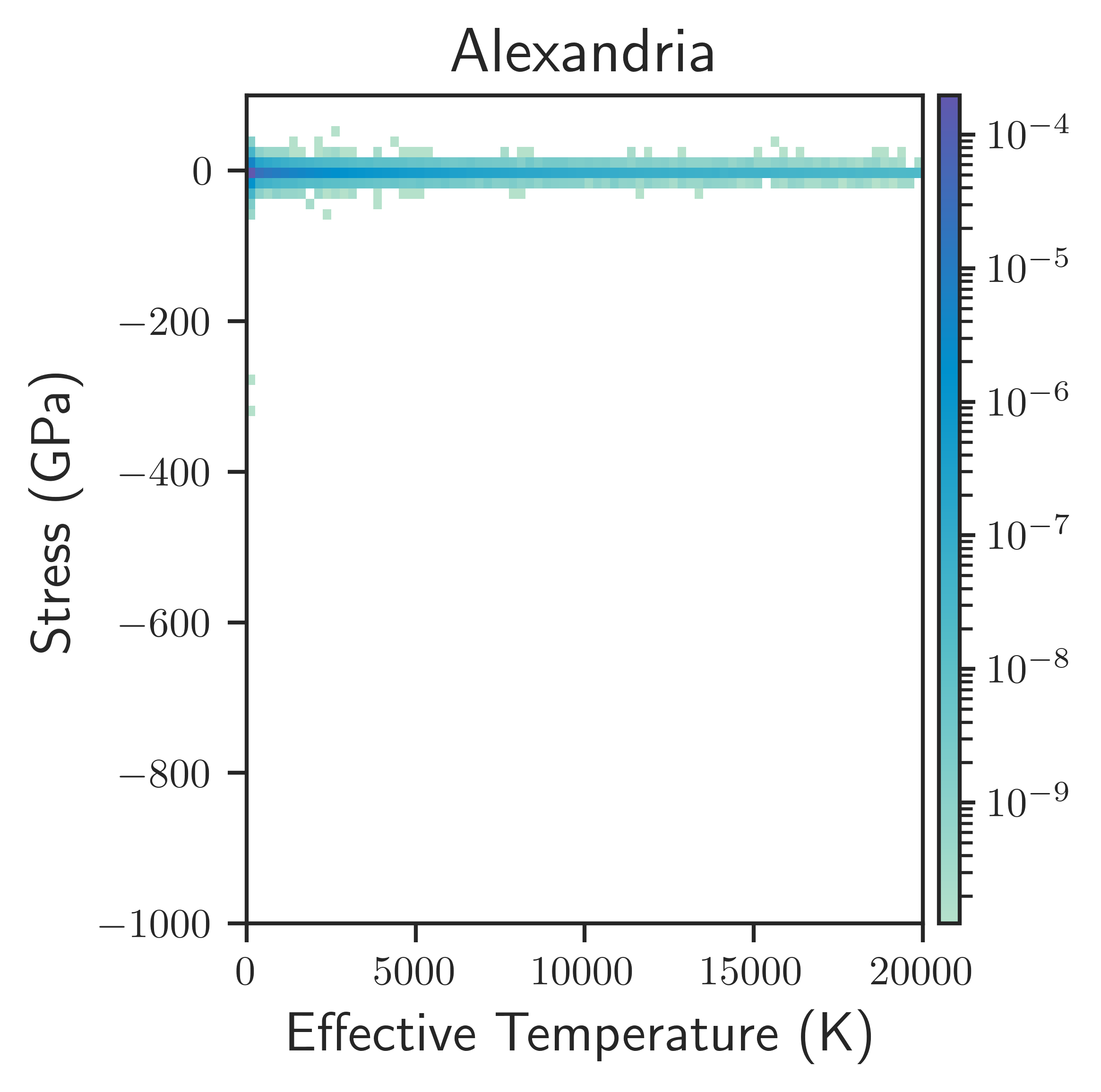}
    \caption{}
    \end{subfigure}
    \begin{subfigure}{0.45\textwidth}
    \includegraphics[width=\textwidth]{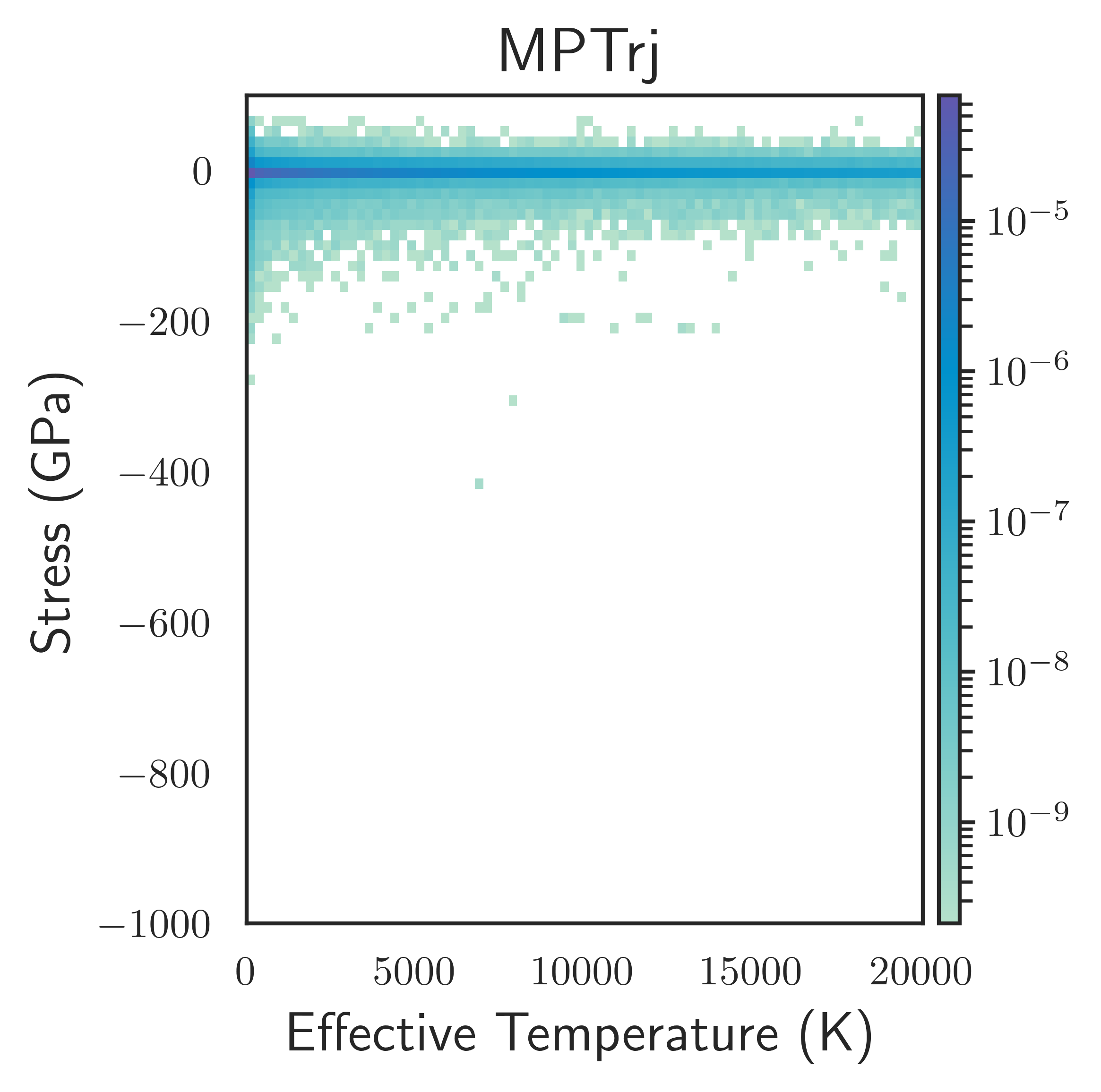}
    \caption{}
    \end{subfigure}
    \begin{subfigure}{0.45\textwidth}
    \includegraphics[width=\textwidth]{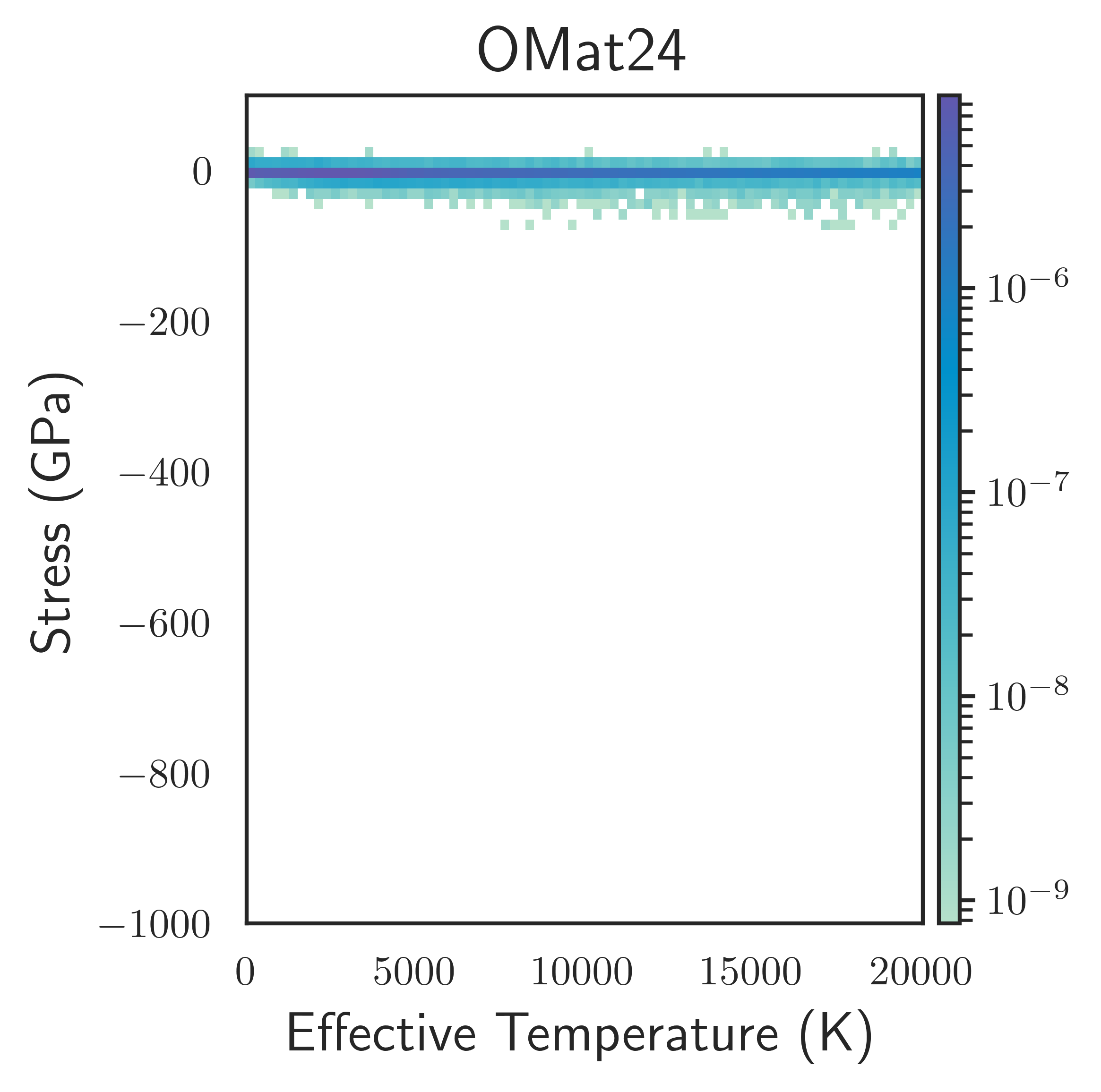}
    \caption{}
    \end{subfigure}
    \begin{subfigure}{0.45\textwidth}
    \includegraphics[width=\textwidth]{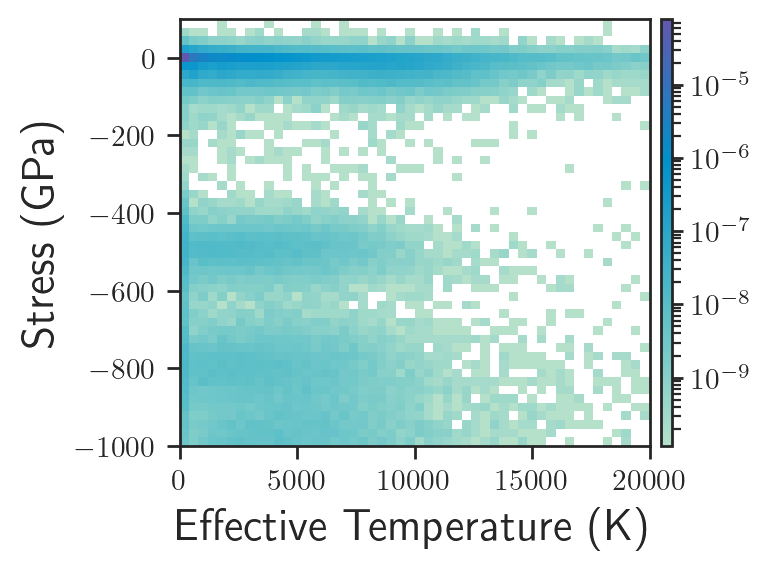}
    \caption{}
    \end{subfigure}
    \begin{subfigure}{0.45\textwidth}
    \includegraphics[width=\textwidth]{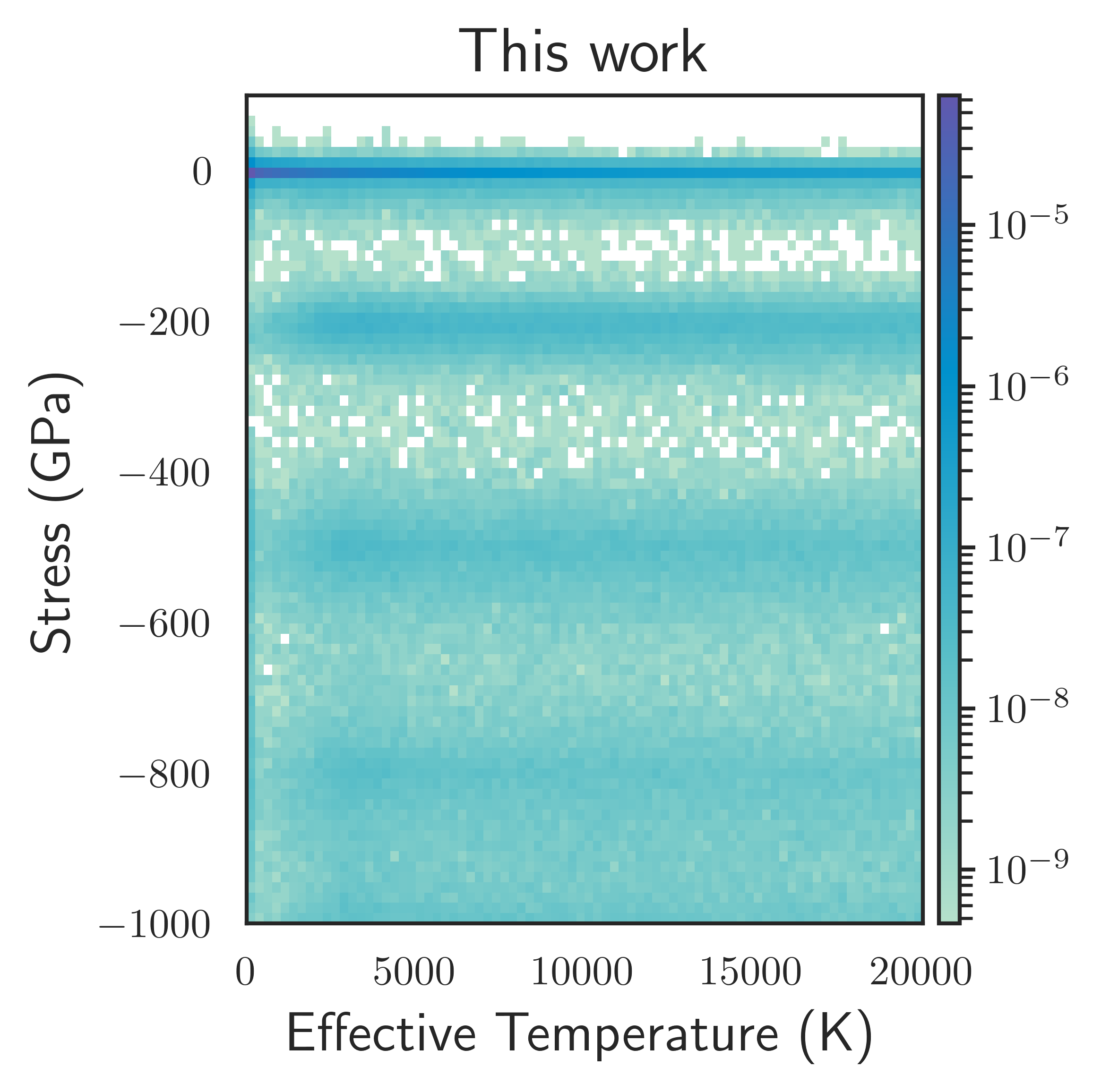}
    \caption{}
    \end{subfigure}
    \caption{Pressure and effective temperature distribution of (a) the entire MPF2021 dataset; and 1M materials randomly sampled from (b) \texttt{Alexandria}, (c) \texttt{MPTrj}, (d) \texttt{OMat24} dataset, (e) MatterSim-v1\cite{yang2024mattersim} and (f) this work. }
    \label{fig:tp-distribution-of-public-dataset}
\end{figure}

The dataset used to train MatterSim contains 35M structures labeled with energies, forces, and stresses, 172,488 with Bader charges, 3,051 with Born effective charges and dielectric matrices, and 284,195 with magnetic moments under ferromagnetic ordering. To straightforwardly illustrate the distribution, we have included a two-dimensional histogram of the effective temperature and stress of the main energetics-related dataset in Fig.~\ref{fig:tp-distribution-of-public-dataset}, where the effective temperature is defined as follows,
\begin{enumerate}
    \item For each material from a given dataset, we evaluate its total energy per atom ($\varepsilon$) with MatterSim;
    \item Then, we optimize the atomic positions with fixed lattice parameters for at most 500 steps until the max forces converge to \SI{0.01}{eV\per \angstrom}, and evaluate the relaxed total energy per atom ($\varepsilon_0$);
    \item Finally, the effective temperature of this given material is evaluated by
    \begin{equation*}
        T_\mathrm{eff} = \frac{\varepsilon-\varepsilon_0}{k_\mathrm{B}},
    \end{equation*}
    where $k_\mathrm{B}$ is the Boltzmann's constant.
\end{enumerate}
With the effective temperature, we compare the distribution of the \texttt{MPF2021} dataset and 1M randomly sampled structures from \texttt{Alexandria}, \texttt{MPTrj} and \texttt{OMat24}, as shown in \autoref{fig:tp-distribution-of-public-dataset}. Since the structures are relaxed to their corresponding local minima, it is not surprising to find that they are densely packed around the \SI{0}{\giga\pascal} in stress, with very scattered data points of high effective temperature and high pressure. The dataset generated in this work, however, has a much wider coverage over the effective temperature space (0 -- \SI{2e4} {\kelvin}) and stress space (0 -- \SI{1000}{\giga\pascal} in magnitude).

We note that the effective temperature should not be directly interpreted as the physical temperature or the temperature employed in the simulations; instead, it is an intuitive metric to measure the energy distribution of the dataset.

\begin{figure}
    \centering
    \includegraphics[width=\linewidth]{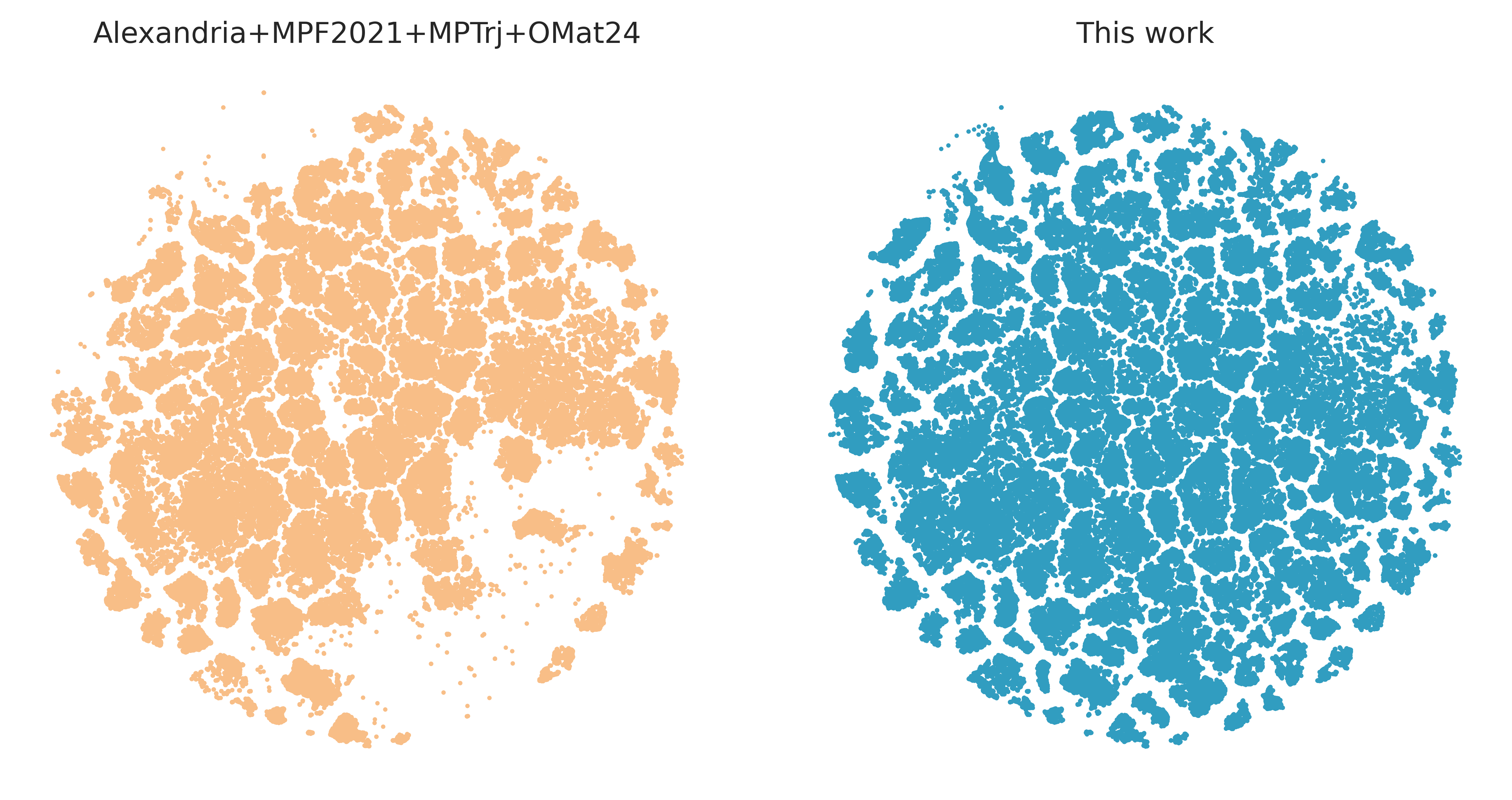}
    \caption{t-SNE analysis of atomic embeddings from the datasets in this work and the comparison with the combination of the Alexandria, MPF2021, MPTrj, and OMat24 datasets. }
    \label{fig:tsne-89elements}
\end{figure}

In addition to comparing the overall temperature and pressure distributions, we analyzed the atomic embeddings to further characterize the configurational space covered in this work. As shown in \autoref{fig:tsne-89elements}, we randomly selected 4,000 atomic embeddings for each element from our dataset, and 1,000 atomic embeddings from each of the MPF2021, MPtrj, Alexandria, and OMat24 datasets, which were then combined for comparison. In \autoref{ms-fig:overview}(b), we performed the same analysis but restricted it to the first 20 elements to provide a clearer view of configurational coverage. Both analyses demonstrate that the dataset presented in this work not only spans the configurational space covered by the four existing datasets, but also substantially extends beyond their combined coverage.

\subsection{Bader charges}
The Bader charge data are retrieved from the AFLOW\cite{curtarolo2013high} database with the
AFLUX API. In total, the data contains the atomic net charges of 172,488 periodic structures. The atomic net charges are the number of valence electrons associated with each atom after Bader decomposition and subtraction of the number of valence electrons in the pseudopotential. The sum of all net atomic charges is zero for every structure.

\subsection{Born effective charges and dielectric matrices}
Born effective charges and dielectric matrices were sourced from PhononDB.\cite{PhononDB} After systematic screening to eliminate structures exhibiting dynamical instabilities (characterized by imaginary frequencies), the final dataset comprises 3,051 viable periodic structures.

\subsection{Magnetic moments}
The magnetic moments are extracted from the MPTrj dataset and are computed under ferromagnetic ordering.\cite{deng2023chgnet}

\section{Benchmarks}\label{sec:benchmarks}
The summary of the out-of-box performance of \texttt{MatterSim-MT-10M} on selected tasks is shown in \autoref{ms-fig:overview}(d). In this section, we describe in more detail the benchmark tasks, datasets used for the benchmarks, and the computational settings employed in the prediction or simulation.

\subsection{Benchmark datasets and results}\label{sec:performance-on-benchmark-sets}
The out-of-box performance of MatterSim is benchmarked on a few datasets computed using the same level of DFT as the training data of MatterSim, including
\begin{itemize}
    \item In-house generated benchmark datasets inherited from \texttt{MatterSim-v1}:
    \begin{itemize}
        \item \texttt{MPF-Alkali-TP}
        \item \texttt{MPF-TP}
        \item \texttt{Random-TP}
        \item \texttt{Extended-TP}
    \end{itemize}
    \item Public dataset
    \begin{itemize}
        \item High-pressure Elemental Xstals (\texttt{HEX}) \cite{giannessi2024database}
    \end{itemize}
\end{itemize}

The \texttt{MPF-Alkali-TP}, \texttt{MPF-TP}, \texttt{Random-TP} benchmark sets are created with increasing complexity to evaluate the models' performance on materials under finite temperature and pressure conditions with far-from-equilibrium atomic positions. All of these benchmark sets are created from first-principles molecular dynamics trajectories initialized with corresponding structures. The \texttt{MPF-Alkali-TP} dataset is sampled from AIMD trajectories of materials that contain alkali metals in Materials Project, and this dataset serves to assess the performance of the model for predicting ionic conductors. The selection rule of elements is that the compound should contain at least one alkali metal and at least one element from N, O, P, S, Se, F, Cl, Br, I. In total, 50 compounds are selected randomly from Materials Project following the selection rule. Similarly, \texttt{MPF-TP} contains molecular dynamics trajectories on 50 randomly selected compounds from the \texttt{MPF2021}\cite{chen2022universal} dataset without elemental constraints. For \texttt{Random-TP}, the initial structures are created by randomly placing 20 atoms with random elements in a simulation box. Again, 50 random structures are used for subsequent molecular dynamics simulation. During collection of the molecular dynamics simulation trajectories, each starting compound is first relaxed followed by running an NPT simulation using VASP, in which the cutoff energy of the plane wave is set to 520 eV and only the gamma point is sampled in reciprocal space to ensure the speed of sampling. The simulation is carried out for 100 ps for each material, and during the last 20 ps, 5 frames are uniformly collected for VASP calculation under MPRelaxSet settings, which are used in the final benchmark set. The temperatures and pressures are all randomly sampled. For the temperature, it is uniformly sampled between 0 and \SI{5000}{\kelvin}. For the pressure, we use a log scale when carrying out the sampling. The pressure range is from 0 to \SI{1000}{\giga\pascal}. These three datasets are designed to test the accuracy of emulators for finite-temperature and pressure simulations with increasing difficulty in generalizability from simple ionic compounds to complex random hypothetical structures. Since the temperature and pressure ranges are wide, these benchmark sets are also reflective of model performance on crystalline materials, amorphous materials, liquids, and pressurized materials.
The \texttt{Extended-TP} dataset is generated in the same way as \texttt{MPF-TP} but with the base structures derived from beyond Materials Project. The used base structures include Materials Project, Alexandria, the relaxed structures from random structure search, and the relaxed structures from MatterGen. This dataset represents the widest coverage of reasonable materials under all temperatures and pressures in this study.

In addition, we also evaluated the performance of MatterSim on the publicly available High-pressure Elemental Xstals (HEX) database,\cite{giannessi2024database} which covers the ground-state structures of elemental crystals for the first 57 elements in the periodic table under \SIrange{0}{300}{\giga\pascal}.

To evaluate the performance, the per-atom mean absolute energy error, the mean error on forces, and mean error on stress are computed for each benchmark set. The results are shown in \autoref{tab:performance-compare}, where the comparison is carried out between MatterSim and a few open-source universal MLFFs, including \texttt{M3GNet}\cite{chen2022universal}, \texttt{CHGNet}\cite{deng2023chgnet}, \texttt{MACE-MP-0}\cite{batatia2023foundation}, \texttt{SevenNet},\cite{park2024scalable} \texttt{Orb-v2},\cite{neumann2024orb} \texttt{OMat24}.\cite{barroso2024open}

Here, we list the checkpoint of each model used in this benchmark:
\begin{itemize}
    \item \texttt{M3GNet}:     \texttt{M3GNet-MP-2021.2.8-PES};
    \item \texttt{CHGNet}: \href{https://github.com/CederGroupHub/chgnet/blob/84e8d55132b2242fad06f9b7f7706b35c1a7da7e/chgnet/pretrained/0.3.0/chgnet_0.3.0_e29f68s314m37.pth.tar}{\texttt{0.3.0}};
    \item \texttt{MACE-MP-0}: \texttt{large} version of the model defined in the commit \texttt{4d2d1c4} in the \href{https://github.com/ACEsuit/mace}{repo};
    \item \texttt{SevenNet}: \texttt{SevenNet-0\_11July2024};
    \item \texttt{Orb-V2}: \href{https://storage.googleapis.com/orbitalmaterials-public-models/forcefields/orb-v2-20241011.ckpt}{\texttt{orb-v2-20241011.ckpt}};
    \item \texttt{OMat24}: \texttt{eqV2-M OMat MPtrj-sAlex}.
\end{itemize}

\begin{table}
    \footnotesize
    \setlength\tabcolsep{2pt}
    \centerline{
    \begin{tabular}{ccccccccc}
    \toprule
    Test Set                  & MAE & OMat24 & ORB-V2 & SevenNet & M3GNet & CHGNet & MACE-MP-0 & MatterSim-MT-10M \\
    \midrule
    \multicolumn{9}{l}{\textbf{In-house generated benchmark datasets}} \\
    \midrule
    \multirow{3}{*}{MPF-Alkali-TP}  & Energy         &         \textbf{0.040} &                    0.142 &                   0.258 &                   0.165 &                   0.250 &                     1.351  & 0.045  \\
                                    & Force          &         \textbf{0.157} &                    0.458 &                   4.495 &                   1.139 &                   1.636 &                     15.819 & 0.488  \\
                                    & Stress         &         \textbf{1.086} &                    6.625 &                  25.039 &                   4.911 &                  12.625 &                     25.723 & 1.252  \\
    \midrule
    \multirow{3}{*}{MPF-TP}         & Energy         &                  0.126 &                    0.267 &                   0.321 &                   0.207 &                   0.254 &                    256.340 & \textbf{0.042}  \\
                                    & Force          &         \textbf{0.233} &                    0.613 &                   4.686 &                   1.224 &                   3.313 &                   1506.854 & 0.530 \\
                                    & Stress         &                  2.408 &                   10.699 &                  22.174 &                   5.575 &                  25.208 &                    202.093 & \textbf{1.368}  \\
    \midrule
    \multirow{3}{*}{Random-TP}      & Energy         &                  0.491 &                    0.822 &                   0.716 &                   0.537 &                   0.506 &                    9.184   & \textbf{0.259}  \\
                                    & Force          &         \textbf{0.578} &                    1.058 &                   8.170 &                   1.789 &                   3.950 &                    88.327  & 1.200  \\
                                    & Stress         &                  \textbf{2.579} &                    7.254 &                  13.696 &                   3.216 &                   7.230 &                    19.224  & 2.715 \\
    \midrule
    \multirow{3}{*}{Extended-TP}    & Energy         &                 2.883  &                    4.135 &           \SI{4.5e10}{} &                   0.781 &                   4.526 &               \SI{2.6e5}{} & \textbf{0.054} \\
                                    & Force          &                  2.015 &                    3.314 &           \SI{3.0e12}{} &                   3.543 &                   5.972 &               \SI{1.2e6}{} & \textbf{0.780}  \\
                                    & Stress         &                 65.847 &                  136.058 &           \SI{2.9e13}{} &            \SI{4.3e4}{} &                 194.145 &               \SI{3.4e7}{} & \textbf{3.274} \\
    \midrule
    \multicolumn{9}{l}{\textbf{Public dataset}} \\
    \midrule
    \multirow{3}{*}{HEX}            & Energy         &                  0.528 &                    0.532 &                   1.463 &                   0.436 &                   0.972 &                    454.462 & \textbf{0.106} \\
                                    & Force          &         \textbf{0.043} &                    0.074 &                  14.638 &                   0.699 &                   0.196 &                    575.443 & 0.118  \\
                                    & Stress         &                 20.902 &                   37.826 &                 325.500 &                  89.833 &                 166.001 &                   2597.080 & \textbf{8.536} \\
    \bottomrule
    \end{tabular}
    }
    \caption{Performance of \texttt{OMat24}, \texttt{Orb-V2}, \texttt{SevenNet}, \texttt{M3GNet}, \texttt{CHGNet}, \texttt{MACE-MP-0}, and MatterSim on benchmark datasets. The mean absolute errors (MAE) of energy, force, and stress are in \SI{}{eV\per atom}, \SI{}{eV\per \angstrom}, and \SI{}{\giga\pascal}, respectively.}\label{tab:performance-compare}
\end{table}

\subsection{Phonons and group velocities}
We benchmark against Materials Data Repository (MDR) phonon calculation database (also known as PhononDB),\cite{PhononDB} a database of phonon properties derived from first-principles calculations. PhononDB encompasses various materials, each characterized by phonon properties computed using the finite displacement method via the Phonopy software package.\cite{phonopy-phono3py-JPCM, phonopy-phono3py-JPSJ} The force constants for these calculations are obtained through VASP\cite{kresse1996efficiency,kresse1996efficient}. Furthermore, the Perdew-Burke-Ernzerhof for solids (PBEsol) exchange-correlation functional\cite{perdew2008restoring,Togo2024private} is utilized within the DFT framework. MatterSim's performance is rigorously assessed against the entire PhononDB database.

Phonon group velocities are the derivatives of phonon frequencies with respect to phonon wavevectors,
\begin{equation}
    \mathbf{v}_{\mathbf{q}\nu} = \frac{\partial \omega_{\mathbf{q}\nu}}{\partial \mathbf{q}},
\end{equation}
which characterize heat transport in crystals and thus play a significant role in solid state physics.

In \autoref{tab:phonon-free-energy-results}, we list the results of the prediction errors for max phonon frequency and phonon group velocities. For phonon frequency prediction, MatterSim has achieved an accuracy of slightly above \SI{1}{\tera\hertz}. In addition, the phonon group velocities predicted with MatterSim are \qtyrange{3}{5}{}-fold better than those of other models.

\subsubsection{Computational setup}
Here the phonon frequencies and group velocities are computed using the finite displacement method implemented in the \texttt{Phonopy} package and the \texttt{Phono3py} packages, respectively.
Before the phonon calculation, we first relax the structure of the materials with the force tolerance of \SI{0.02}{eV\per\angstrom} for up to 500 steps, which relaxes both the cell size and the atomic positions of the initial structure.
To generate forces required for the phonon and group velocity predictions, displacements compatible with space group are introduced to atomic positions within the supercell, and the size of the supercell is parametrized to be the same as in the PhononDB database. With a magnitude of \SI{0.03}{\angstrom}, consistent with the settings in PhononDB, the forces acting on each displaced atom are then predicted using either MatterSim or other models.

\begin{table}
    \centering
    \footnotesize
    \setlength\tabcolsep{2pt}
    \centerline{
    \begin{tabular}{cccccccccc}
    \toprule
    \multirow{2}{*}{MAE of} & \multirow{2}{*}{OMat24} & \multirow{2}{*}{ORB-V2} & \multirow{2}{*}{SevenNet} & \multirow{2}{*}{M3GNet}& \multirow{2}{*}{CHGNet} & \multirow{2}{*}{MACE-MP-0} & MatterSim   \\
                            &      &       &       &                         &                         &                          & (10M)   \\
    \midrule
    Max phonon frequency (\SI{}{\tera\hertz}) &7.007 &3.060 & 1.412 &4.248 & 2.397 & 1.821 & \textbf{1.016} \\
    Phonon group velocity (\SI{}{\kilo\meter / \second})  &168.386 & 50.959 & 69.663 &110.073 & 109.495 & 39.930 & \textbf{22.895} \\
    $\Delta\Delta G$ vs DFT (\SI{}{meV\per atom})  &28.064 &207.715 &26.959  &47.333 &166.099  &35.433  & \textbf{18.504 }   \\
    \bottomrule
    \end{tabular}
    }
    \caption{Performance of \texttt{OMat24}, \texttt{Orb-V2}, \texttt{SevenNet}, \texttt{M3GNet}, \texttt{CHGNet}, \texttt{MACE-MP-0}, and MatterSim on prediction of max phonon frequency, max phonon group velocity, and Gibbs free energy. }\label{tab:phonon-free-energy-results}
\end{table}

\subsection{Free energy prediction with QHA}
To construct an accurate phase diagram, we calculate the free energies of materials using the quasi-harmonic approximation (QHA) implemented in \texttt{Phonopy},\cite{togo2023implementation} which is a theoretical framework that allows for the calculation of free energies at different temperatures and pressures.

The quasi-harmonic approximation is an extension of the harmonic approximation. The harmonic approximation assumes that atoms in a crystal vibrate about their equilibrium positions and the potential energy can be approximated by a quadratic function of the atomic displacements. This model is accurate at low temperatures where anharmonic effects are negligible. However, as the temperature increases, anharmonic contributions become significant, and the harmonic approximation fails to predict the correct thermodynamic behavior. This is where the QHA becomes valuable.

QHA takes into account the volume dependence of the vibrational frequencies. It assumes that while the shape of the potential energy well changes with volume, each volume can still be described by a harmonic potential, but with different force constants. QHA allows the phonon frequencies to depend on the volume, thereby capturing the thermal expansion effect.

The free energy within the quasi-harmonic approximation is calculated by considering both the vibrational and static lattice contributions. The Helmholtz free energy, $F$, at a given temperature ($T$) and volume ($V$) can be expressed as:
\begin{equation}
    F(T, V) = U(V) - T S(T, V),
\end{equation}
where $U$ is the internal lattice energy and $S$ is the entropy due to the vibrational degrees of freedom.

The entropy term equals
\begin{equation}
    S(T, V) = -\frac{1}{N} \sum_{\mathbf{q} \nu} k_B \ln{\left[ 1 - \exp(-\hbar \omega_{\mathbf{q} \nu}(V) / k_B T) \right]} + \frac{1}{N T} \sum_{\mathbf{q}\nu} \frac{\hbar \omega_{\mathbf{q}\nu}(V)}{\exp(\hbar \omega_{\mathbf{q}\nu}(V) / k_B T) - 1},
\end{equation}
where $\omega_{\mathbf{q}\nu}$ is the $\nu$-th phonon mode at $\mathbf{q}$ wavevector in the Brillouin zone, $\hbar$ is the reduced Planck constant, $k_B$ is the Boltzmann constant and $T$ is the temperature.

The Gibbs free energy $G$ is obtained by
\begin{equation}
    G(T, P) = \min_{V} \left[ U(V) - T S(T, V) + P V \right]
\end{equation}

\subsubsection{Computational setup}\label{sec:qha-computational-setup}
The Gibbs free energy of ordered crystalline materials is computed using MatterSim via the quasiharmonic approximation (QHA) implemented in Phonopy. \autoref{tab:free-energy-benchmark-materials} lists the materials for which the Gibbs free energies are computed.
For each material, we change the volume by -2.5\% to 2.5\% and relax the structure with a force tolerance of \SI{0.02}{eV\per \angstrom}. The phonon calculations are performed with a $4\times 4\times 4$ supercell.

For benchmark, we first compute the change in Gibbs free energy $\Delta G$ between \SI{300}{\kelvin} and \SI{900}{\kelvin}, and then compute the difference ($\Delta\Delta G$) between the MatterSim and DFT results. The DFT reference Gibbs free energies are computed at the DFT/PBE level of theory, and the results are listed in \autoref{tab:phonon-free-energy-results}.

\begin{table}
    \centering
    \caption{Summary of materials and their corresponding ID in Materials Project used for the benchmarks of Gibbs free energy prediction.}\label{tab:free-energy-benchmark-materials}
    \begin{tabular}{lc|lc|lc|lc}
    \toprule
    mp-id & formula & mp-id & formula & mp-id & formula & mp-id & formula \\
    \midrule
    mp-1000    & \ce{BaTe}   & mp-1002164 & \ce{CGe}    & mp-10044   & \ce{AsB}    & mp-1007652 & \ce{As2In2} \\
    mp-1007661 & \ce{In2Sb2} & mp-1008786 & \ce{MgTe}   & mp-1017439 & \ce{O3SiSr} & mp-1018059 & \ce{Ga2Sb2} \\
    mp-1018100 & \ce{Al2Sb2} & mp-10250   & \ce{BaF3Li} & mp-1039    & \ce{Mg2Te2} & mp-10695   & \ce{SZn}    \\
    mp-1070    & \ce{Cd2Se2} & mp-10760   & \ce{MgSe}   & mp-1087    & \ce{SSr}    & mp-1156    & \ce{GaSb}   \\
    mp-11718   & \ce{FRb}    & mp-1184046 & \ce{Cl2Cu2} & mp-1190    & \ce{SeZn}   & mp-1253    & \ce{BaSe}   \\
    mp-1265    & \ce{MgO}    & mp-12779   & \ce{Cd2Te2} & mp-13031   & \ce{MgSe}   & mp-13033   & \ce{MgTe}   \\
    mp-1315    & \ce{MgS}    & mp-1342    & \ce{BaO}    & mp-1367    & \ce{Mg2Si}  & mp-1415    & \ce{CaSe}   \\
    mp-1479    & \ce{BP}     & mp-149     & \ce{Si2}    & mp-1500    & \ce{BaS}    & mp-1519    & \ce{CaTe}   \\
    mp-1541    & \ce{BeSe}   & mp-1550    & \ce{AlP}    & mp-1639    & \ce{BN}     & mp-1672    & \ce{CaS}    \\
    mp-1700    & \ce{AlN}    & mp-1778    & \ce{BeO}    & mp-1883    & \ce{SnTe}   & mp-1958    & \ce{SrTe}   \\
    mp-1960    & \ce{Li2O}   & mp-19717   & \ce{PbTe}   & mp-1986    & \ce{OZn}    & mp-20012   & \ce{InSb}   \\
    mp-20194   & \ce{CeO2}   & mp-20305   & \ce{AsIn}   & mp-20351   & \ce{InP}    & mp-20411   & \ce{InN}    \\
    mp-20724   & \ce{Mg2Pb}  & mp-2133    & \ce{O2Zn2}  & mp-2172    & \ce{AlAs}   & mp-2201    & \ce{PbSe}   \\
    mp-22205   & \ce{In2N2}  & mp-22865   & \ce{ClCs}   & mp-22894   & \ce{Ag2I2}  & mp-22905   & \ce{ClLi}   \\
    mp-22906   & \ce{BrCs}   & mp-22913   & \ce{BrCu}   & mp-22914   & \ce{ClCu}   & mp-23193   & \ce{ClK}    \\
    mp-23209   & \ce{Cl2Sr}  & mp-23251   & \ce{BrK}    & mp-23259   & \ce{BrLi}   & mp-241     & \ce{CdF2}   \\
    mp-2469    & \ce{CdS}    & mp-2472    & \ce{OSr}    & mp-2490    & \ce{GaP}    & mp-252     & \ce{BeTe}   \\
    mp-2534    & \ce{AsGa}   & mp-2542    & \ce{Be2O2}  & mp-2597    & \ce{InTe}   & mp-2605    & \ce{CaO}    \\
    mp-2624    & \ce{AlSb}   & mp-2653    & \ce{B2N2}   & mp-2691    & \ce{CdSe}   & mp-2730    & \ce{HgTe}   \\
    mp-2758    & \ce{SeSr}   & mp-32      & \ce{Ge2}    & mp-3448    & \ce{F3KMg}  & mp-380     & \ce{Se2Zn2} \\
    mp-406     & \ce{CdTe}   & mp-408     & \ce{GeMg2}  & mp-422     & \ce{BeS}    & mp-463     & \ce{FK}     \\
    mp-560588  & \ce{S2Zn2}  & mp-561947  & \ce{CsF3Hg} & mp-567636  & \ce{FeSbV}  & mp-568560  & \ce{BrTl}   \\
    mp-569346  & \ce{Cu2I2}  & mp-5878    & \ce{F3KZn}  & mp-5967    & \ce{CoSbTi} & mp-643     & \ce{O2Th}   \\
    mp-66      & \ce{C2}     & mp-661     & \ce{Al2N2}  & mp-672     & \ce{Cd2S2}  & mp-682     & \ce{FNa}    \\
    mp-7104    & \ce{CaCsF3} & mp-7140    & \ce{C2Si2}  & mp-804     & \ce{Ga2N2}  & mp-8062    & \ce{CSi}    \\
    mp-820     & \ce{HgSe}   & mp-830     & \ce{GaN}    & mp-8399    & \ce{CdCsF3} & mp-8402    & \ce{F3MgRb} \\
    mp-8880    & \ce{Al2P2}  & mp-8881    & \ce{Al2As2} & mp-8882    & \ce{Ga2P2}  & mp-8883    & \ce{As2Ga2} \\
    mp-8884    & \ce{Te2Zn2} & mp-924129  & \ce{NiSnZr} & mp-924130  & \ce{NiSnTi} & mp-966800  & \ce{In2P2}  \\
    mp-978489  & \ce{O3PbSi} & mp-997618  & \ce{BSb}    & mp-998739  & \ce{F3MgTl} &            &             \\
    \bottomrule
    \end{tabular}
\end{table}

\subsection{Results of Multiple Tasks}
The results of different tasks on the test set are shown in \autoref{tab:multi-task-results}.
\begin{table}
\centering
\caption{Multi-task Result}
\begin{tabular}{lc}
\toprule
\textbf{Multi-task Valid MAE}  \\
\midrule
magnetic moments ($\mu_B$)                   & 0.0640    \\
Bader charges ($e$)                       & 0.0233 \\
dielectric matrices               & 0.2478 \\
Born effective charges ($e$)           & 0.0756 \\
\bottomrule
\end{tabular}
\label{tab:multi-task-results}
\end{table}

\section{Multi-task Case studies}\label{sec:case-study}

\subsection{Formation enthalpy of \ce{LaH10}}\label{sec:case-study-LaH10}
The phase stability of the La--H system is investigated by constructing convex hulls based on the enthalpy of potential phases under pressures ranging from \qtyrange{0}{200}{\giga\pascal}. All calculations are performed using MatterSim-MT \SI{10}{M} parameters. The analysis involves 21 distinct phases:
\begin{itemize}
    \item seven La-H phases from the Materials Project with the MP IDs: \texttt{mp-1104116}, \texttt{mp-1103808}, \texttt{mp-24153}, \texttt{mp-1018144}, \texttt{mp-1103947}, \texttt{mp-973064}, \texttt{mp-2647067};
    \item four ground-state Hydrogen phases under 0, 100, 200, and \SI{300}{\giga\pascal} from Ref.~\citenum{giannessi2024database};
    \item four ground-state Lanthanum phases under 0, 100, 200, and \SI{300}{\giga\pascal} from  Ref.~\citenum{giannessi2024database};
    \item six high-pressure La-H phases identified through random structure search \ce{LaH10} ($\mathrm{Fm\bar{3}m}$), \ce{LaH3} ($\mathrm{Cmcm}$), \ce{LaH5} ($\mathrm{P\bar{I}}$), \ce{LaH2} ($\mathrm{C2/m}$), \ce{LaH4} ($\mathrm{I4/mmm}$), and \ce{LaH8} ($\mathrm{C2/m}$).
\end{itemize}
To determine the pressure-dependent enthalpy of each phase, structures are initially relaxed at \SI{200}{\giga\pascal}, allowing optimization of both atomic coordinates and cell parameters. Subsequent relaxations are performed with decreasing pressure at \SI{20}{\giga\pascal} intervals down to \SI{0}{\giga\pascal}. The enthalpy is computed following each relaxation. Convex hulls are constructed using the \texttt{PhaseDiagram} package from \texttt{pymatgen}\cite{ong2013python} with enthalpies normalized such that the lowest-enthalpy end members (H and La) are set to zero.

\begin{figure}
    \centering
    \includegraphics[width=\linewidth]{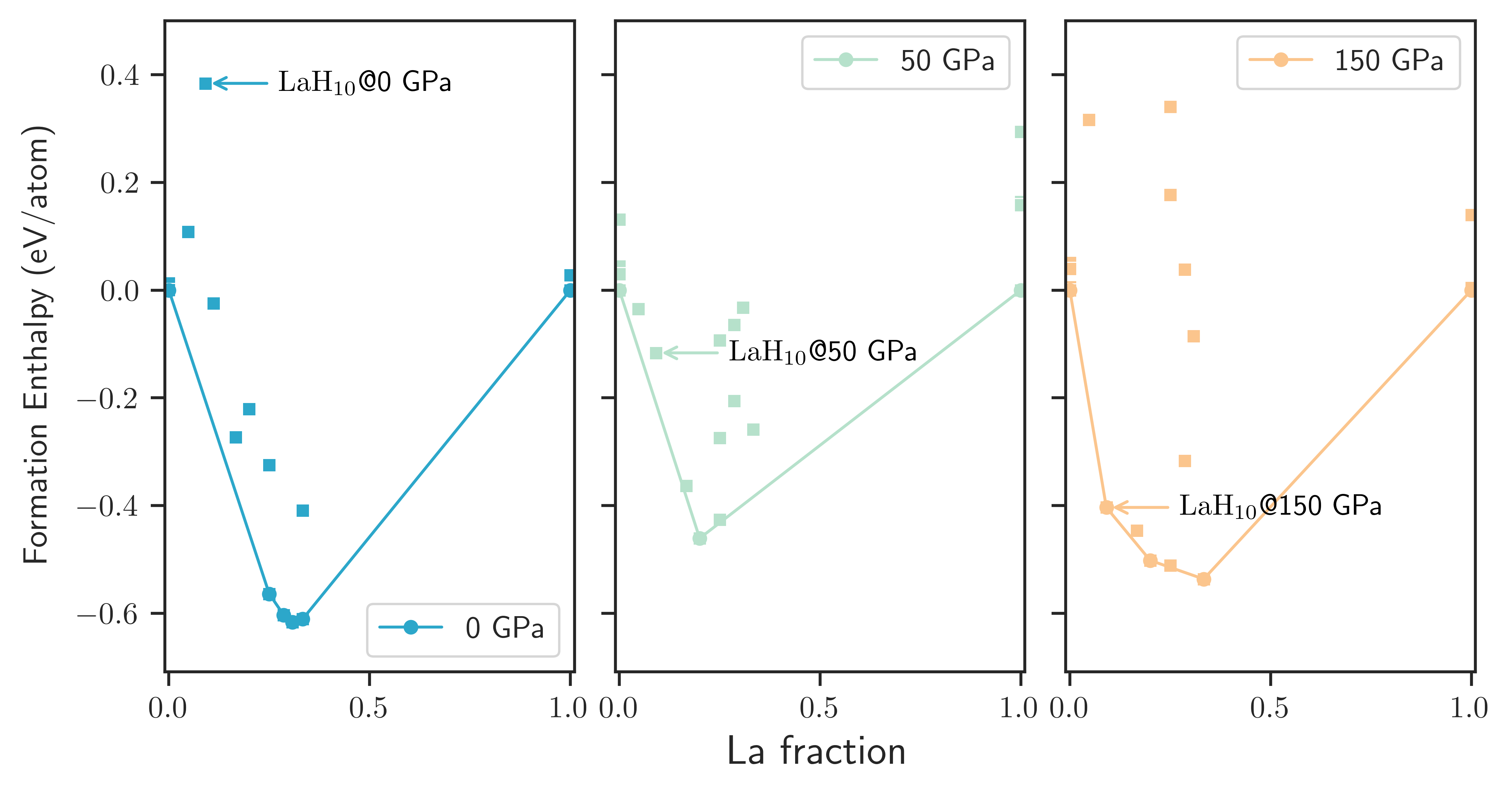}
    \caption{The formation enthalpy of the La--H chemical system under the pressure of 0, 50, and \SI{150}{\giga\pascal}. }
    \label{fig:case-study-LaH10}
\end{figure}

To validate the MatterSim results, density functional theory (DFT) calculations are performed using identical protocols at selected pressures: 0, 50, and \SI{150}{\giga\pascal}, as presented in \autoref{fig:case-study-LaH10}.
The hydrogen-rich phase \ce{LaH10} exhibits pressure-dependent stability: it is thermodynamically unstable at ambient pressure but becomes stabilized at \SI{150}{\giga\pascal}, consistent with the MatterSim predictions shown in \autoref{ms-fig:hpht}(a). This computational finding aligns with previous first-principles simulations that predict \ce{LaH10} stability at \SI{150}{\giga\pascal}\cite{kruglov2020superconductivity} and experimental observations of \ce{LaH10} formation within a pressure range of \qtyrange{137}{218}{\giga\pascal}\cite{drozdov2019superconductivity}.

\subsection{Pressure dependent LO--TO splitting of $3c$-\ce{SiC}}\label{sec:case-study-sic}
Due to its high thermal conductivity and wide band gap, $3c$-\ce{SiC} is believed to have potential applications in the semiconductor industry.
As a polar material, $3c$-\ce{SiC} has a finite LO--TO splitting that cannot be predicted using traditional force fields because the prediction of Born effective charges and dielectric matrices is required for this purpose.

To showcase the multi-task capability of MatterSim, we predict the pressure-dependent LO--TO splitting of $3c$-\ce{SiC} as illustrated in \autoref{ms-fig:multi-task-capabilities}(b), by adding the non-analytical correction to the force constant matrices:\cite{cochran1962dielectric,giannozzi1991ab,baroni2001phonons}
\begin{equation}
^\text{NA}C_{I\alpha,J\beta} = \frac{4\pi}{\Omega} \frac{(\mathbf{q}\cdot Z^*_I)_\alpha(\mathbf{q} \cdot Z^*_J)_\beta}{\mathbf{q}\cdot \mathbf{\varepsilon}_\infty \cdot \mathbf{q}},
\end{equation}
where $Z^*_I$ is the Born effective charge on the $I$-th atom, $\mathbf{\varepsilon}_\infty$ is the dielectric matrix, $\alpha$, $\beta$, $\gamma$ are indices of three dimensions, and $\mathbf{q}$ is the wavevector of the phonon.

In \autoref{tab:sic-bec-and-dielectric}, we list the lattice constant $a$, Born effective charge on Silicon $Z_\text{Si}^*$, dielectric constant $\varepsilon_\infty$, longitudinal optical phonon frequency at $\Gamma$ point $\omega_\text{LO}(\Gamma)$, transverse optical phonon frequency at $\Gamma$ point $\omega_\text{TO}(\Gamma)$ and the LO--TO splitting predicted with MatterSim of $3c$-\ce{SiC} at zero pressure. MatterSim predicts the LO-TO splitting with an error of only \SI{0.06}{\tera\hertz} with respect to previous literature employing \emph{ab initio} computational methods\cite{karch1996pressure}, and \SI{0.03}{\tera\hertz} when compared to experimental measurements.\cite{olego1982pressure-frequency}
\begin{table}
    \centering
    \caption{Lattice constant $a$, Born effective charge on Silicon $Z_\text{Si}^*$, dielectric constant $\varepsilon_\infty$, longitudinal optical phonon frequency at $\Gamma$ point $\omega_\text{LO}(\Gamma)$, transverse optical phonon frequency at $\Gamma$ point $\omega_\text{TO}(\Gamma)$ and LO--TO splitting predicted with MatterSim of $3c$-\ce{SiC} at zero pressure in comparison with literature.}
    \begin{tabular}{cccc}
    \toprule
         & MatterSim & Theoretical & Experimental \\
    \midrule
    $a$ (\AA)    & 4.377   & 4.385 (PBE)\cite{wang2010surface}, 4.38 (PBE)\cite{skone2014self} & 4.360\cite{madelung1982physics} \\
    $Z^*_\text{Si}$ (a.u.) & 2.71 & 2.72(LDA)\cite{karch1996pressure} & 2.70\cite{olego1982pressure-BEC} \\
    $\varepsilon_\infty$ & 7.31 & 7.02(LDA)\cite{karch1996pressure}, 6.94 (PBE)\cite{skone2014self} & 6.52\cite{yu2005fundamentals} \\
    $\omega_\text{LO}(\Gamma)$ (\SI{}{\tera\hertz}) & 28.54 & 28.56(LDA)\cite{karch1996pressure} & 29.16\cite{olego1982pressure-frequency} \\
    $\omega_\text{TO}(\Gamma)$ (\SI{}{\tera\hertz}) & 23.28 & 23.36(LDA)\cite{karch1996pressure} & 23.87\cite{olego1982pressure-frequency} \\
    LO--TO splitting (\SI{}{\tera\hertz}) & 5.26 & 5.20 (LDA)\cite{karch1996pressure} & 5.29\cite{olego1982pressure-frequency} \\
    \bottomrule
    \end{tabular}
    \label{tab:sic-bec-and-dielectric}
\end{table}

\subsection{Phase boundary of B1--B2 \ce{MgO}}\label{sec:case-study-mgo}
To construct an accurate phase boundary of the B1--B2 phases of \ce{MgO}, we calculate the free energies of the two phases using the QHA methodology as implemented in \texttt{Phonopy},\cite{togo2023implementation}. The computational setup we employ is the same as in \autoref{sec:qha-computational-setup}.

\subsection{Hysteresis curve of \ce{BaTiO3}}

The simulation of \ce{BaTiO3} under an applied electric field is based on the electric enthalpy,
\begin{equation}
    U = U_0 - \mathbf{E}\cdot \mathbf{P},
\end{equation}
where $U_0$ is the potential energy in the absence of an electric field, $\mathbf{E}$ is the applied electric field, and $\mathbf{P}$ is the polarization of the material. Using the Born effective charges predicted by MatterSim-MT, we evaluate the polarization of unit cell $u$ as
\begin{equation}
    \mathbf{P}^{(u)} = \frac{1}{2} \sum_{O\in u} \mathbf{Z}_\mathrm{O}^* \cdot \Delta r_\mathrm{O-Ti}^{(u)}
    + \frac{1}{8} \sum_{\mathrm{Ba}\in u} \mathbf{Z}_\mathrm{Ba}^* \cdot \Delta r_\mathrm{Ba-Ti}^{(u)},
\end{equation}
where $\Delta r$ denotes the relative displacement between atoms, and $\mathbf{Z}^*$ is the Born effective charge tensor. The applied electric field also introduces an additional force on each atom during the MD simulation:
\begin{equation}
    F_i = F_i^0 + e \mathbf{Z}_i^* \cdot \mathbf{E},
\end{equation}
where $F_i^0$ is the force in the absence of the electric field.

With this setup, we performed MD simulations at target temperatures of 100, 200, and 300 K using a Nos\'e--Hoover thermostat with time step \SI{2}{fs} under a sinusoidal electric field,
\begin{equation}
    \mathbf{E}(t) = E_\mathrm{max} \cos\left(\frac{2\pi t}{T}\right),
\end{equation}
where the maximum electric field is $E_\mathrm{max} = \SI{4}{MV\per\centi\meter}$ and the period is $T = \SI{200}{\pico\second}$. The hysteresis curve shown in \autoref{ms-fig:multi-task-capabilities}(c) is averaged over 5 cycles.

\subsection{Delithiation of \ce{Li_{1.2-x}Mn_{0.8}O2}}
The molecular dynamics simulations of \ce{Li_{1.2-x}Mn_{0.8}O2} are conducted using the MatterSim-MT-10M model, following protocols similar to those in Ref.~\citenum{mccoll2024phase}.
For the simulation, we construct an initial structure of \ce{Li_{1.2-x}Mn_{0.8}O2}  by generating a $5 \times 4 \times 2$ supercell from the conventional cell of \ce{LiCoO2} (mp-22526). Subsequently, all Co atoms are replaced with Mn, and 30 Mn atoms are randomly selected and substituted with Li. During the simulation, the structure undergoes initial relaxation on both ionic and lattice degrees of freedom. Molecular dynamics simulations are then performed under the NVT ensemble at \SI{1000}{\kelvin} with a \SI{1}{\femto\second} timestep. One lithium atom is removed every \SI{5}{\pico\second} until complete delithiation, resulting in a total simulation time of approximately \SI{900}{\pico\second}. In addition to forces, energies, and stress tensors, Bader charges and atomic magnetic moments are computed using MatterSim for real-time analysis.

\begin{figure}
    \centering
    \includegraphics[width=\linewidth]{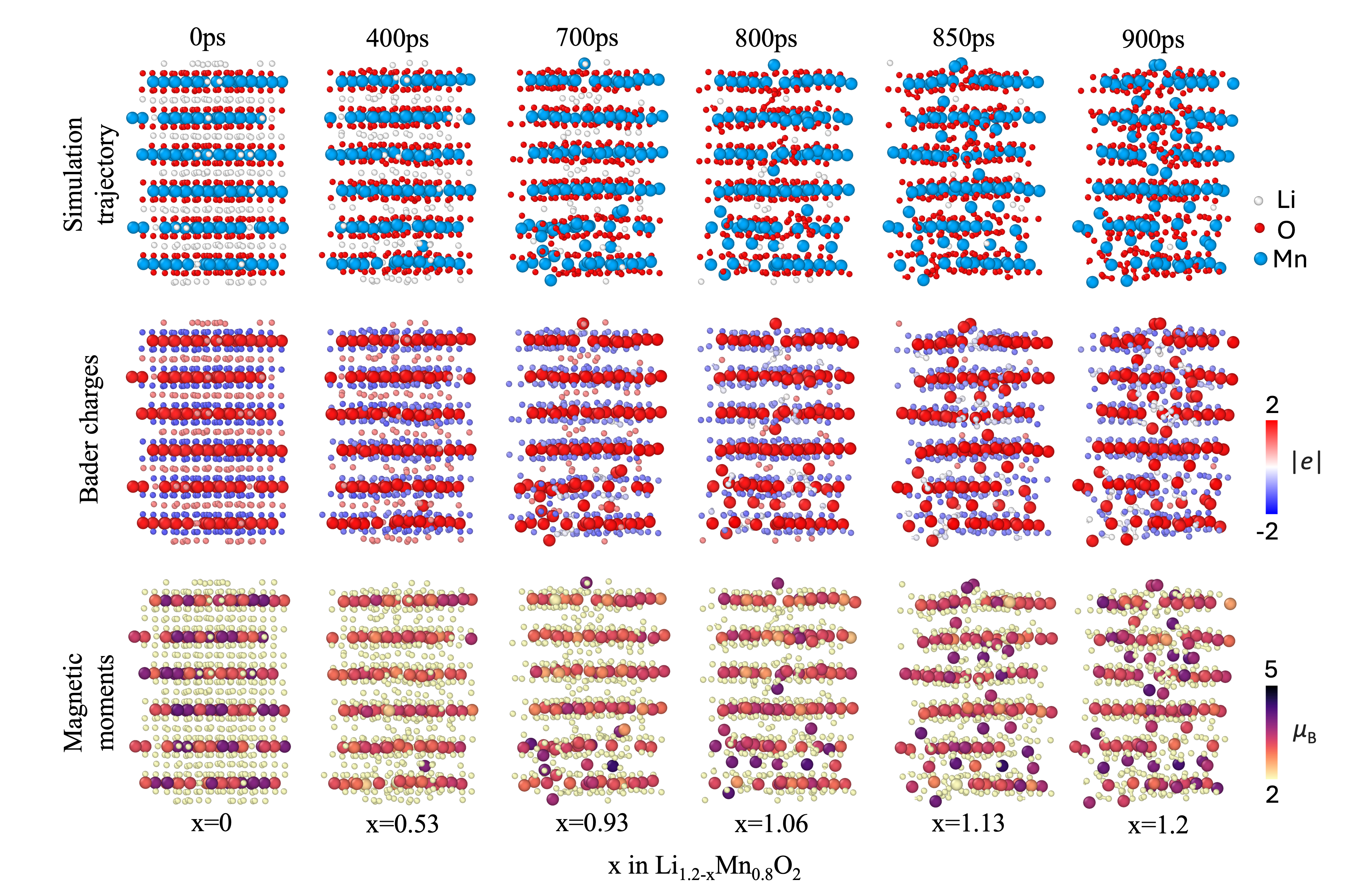}
    \caption{Snapshots of multi-task molecular dynamics simulation on \ce{Li_{1.2-x}Mn_{0.8}O2}}
    \label{fig:case-study-li-battery}
\end{figure}

\autoref{fig:case-study-li-battery} illustrates the structural evolution at various lithium concentrations. Initially, Mn acts as the primary redox center, characterized by a change in magnetic moment from approximately \qtyrange{3}{4}{\mu_B}. When lithium depletion reaches $x \approx 0.5$ in \ce{Li_{1.2-x}Mn_{0.8}O2}, Mn ions begin migrating from octahedral to tetrahedral sites in the lithium layers, accompanied by oxygen sublattice distortions. At this stage, most oxygen atoms maintain their anionic state. Further delithiation to $x \approx 0.9$ triggers oxygen dimer formation, typically following Mn migration to the Li layer. The evolution of these unstable dimers leads to molecular oxygen formation within the lattice, signifying irreversible anionic redox. This phenomenon is corroborated by Bader charge analysis, where the \ce{O2} molecules exhibit near-zero charges. Continued delithiation destabilizes the lattice structure, resulting in increased molecular oxygen formation and eventual transition to nanofluidic states.

\subsection{Comparison to other foundational interatomic potentials}

To put the performance of \texttt{MatterSim-MT} into context, we assess how well popular and established ML interatomic potentials capture the energetics relevant to high-pressure
phase stability.
Specifically, we benchmark several publicly available models including MACE-OMAT-0~\cite{batatia2023foundation}, PET OAM XL~\cite{}, ORB v3~\cite{}, eSEN-30M-OAM~\cite{}, and MatterSim v1~\cite{yang2024mattersim} based on the M3GNet  architecture~\cite{chen2022universal}, on two tasks from the main text: the La–H
convex hull under pressure and the MgO B1–B2 phase boundary.

\begin{figure}
    \centering
    \includegraphics[width=\linewidth]{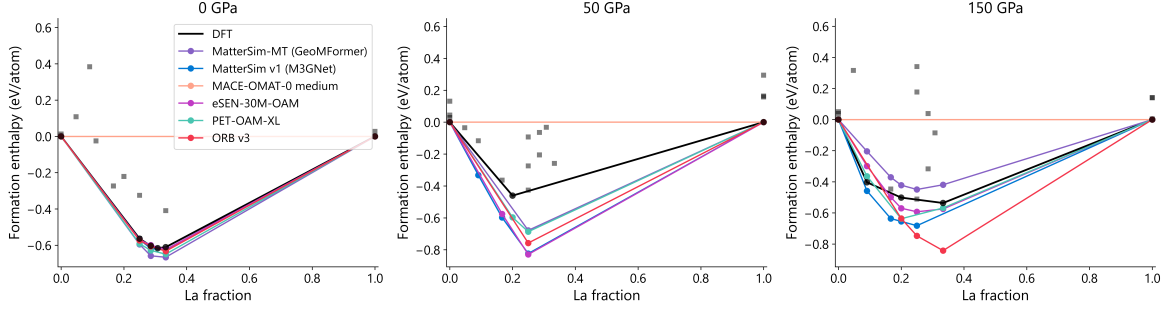}
    \caption{ Convex hull of formation enthalpies in the La–H system at 0, 50, and 150 GPa. For clarity, we only show
  for DFT the phases above the hull (black squares). Coloured lines show the convex hulls predicted by ML
  interatomic potentials.}
    \label{fig:case-study-bench-LaH-hulls}
\end{figure}

We begin with the La–H system.
For each model, we follow the workflow described in section \ref{sec:case-study-LaH10} and relax all structures at pressures between 200 and 0 GPa with a force tolerance of 0.02 eV/Å.
In Fig.~\ref{fig:case-study-bench-LaH-hulls}, we construct the convex hull of formation enthalpies for each model at 0, 50, and 150 GPa and compare the results to the DFT calculations, reported in Fig.~\ref{fig:case-study-LaH10}.
At ambient pressure, all models apart from MACE-OMAT-0 medium reproduce the DFT hull within $\approx$50,meV/atom, correctly identifying LaH$_2$, LaH$_3$, and La$_2$H$_5$ as stable phases.
MACE-OMAT-0 medium produces unphysical relaxed geometries at elevated pressures that propagate to the ambient-pressure hull through the
chained relaxation protocol.
A reverse sweep from 0 to 200~GPa confirms that the model reproduces the DFT hull at 0~GPa but diverges at pressures as low as 50~GPa.
At 50~GPa, the hull topology is sensitive to small energy differences and varies across models.
DFT places only LaH$_4$ on the hull, whereas most models also stabilise LaH$_3$ and some LaH$_5$, which lie only $\approx$6 and $\approx$21,meV/atom above the DFT hull, respectively.
At 150~GPa, all models except MACE-OMAT-0 medium and ORB v3 reproduce the DFT hull, including the stabilisation of LaH$_{10}$, LaH$_4$, and LaH$_2$. Similar to 50~GPa, some models place additional phases (LaH$_3$, LaH$_5$) on the hull which are marginally above the DFT hull.

\begin{figure}
    \centering
\includegraphics[width=0.5\linewidth]{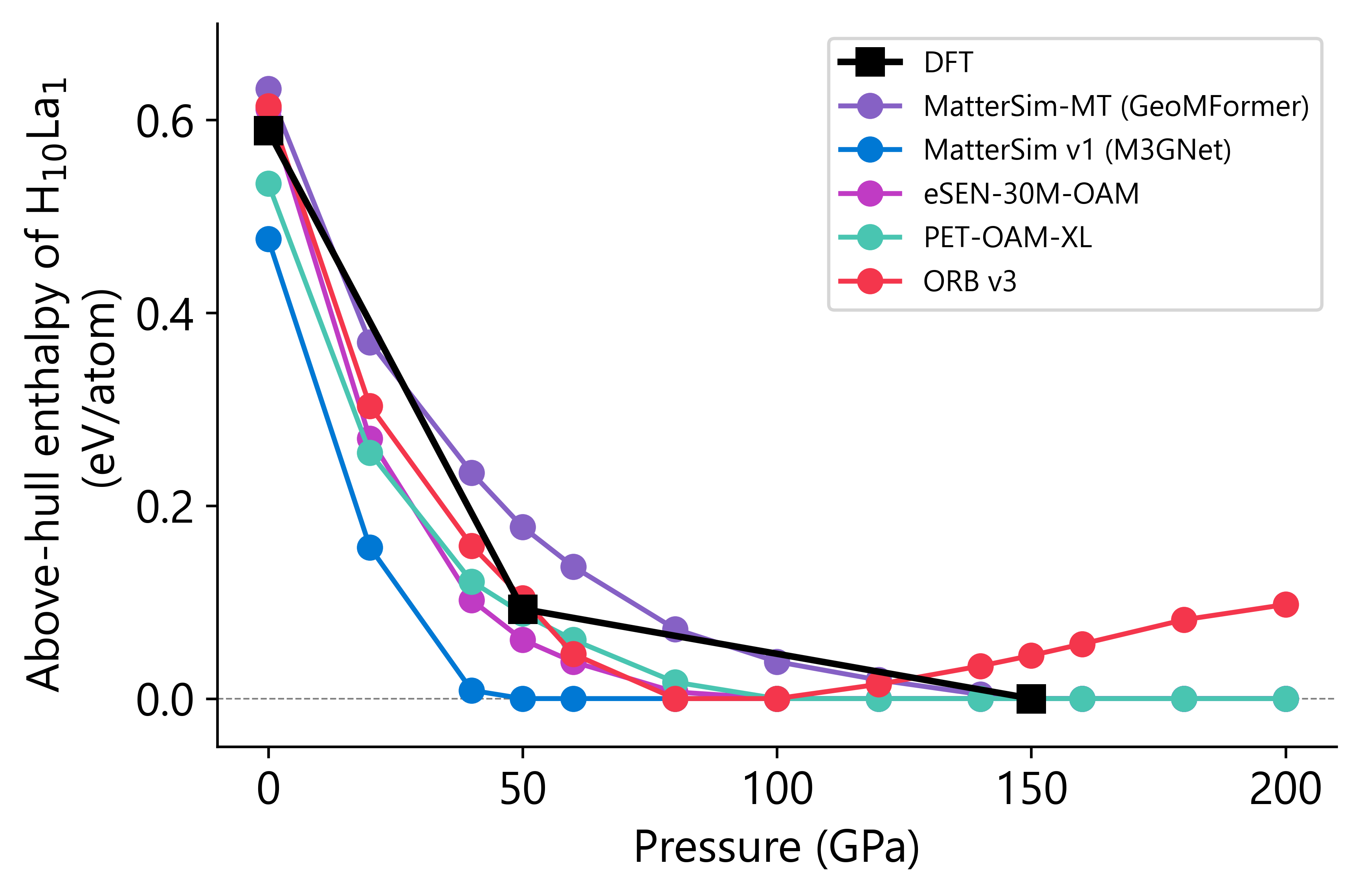}
    \caption{Above-hull enthalpy of LaH$_{10}$ as a function of pressure predicted by DFT and a range of ML interatomic potentials.}
    \label{fig:case-study-bench-LaH10}
\end{figure}

Fig.~\ref{fig:case-study-LaH10} tracks the above-hull enthalpy of LaH${10}$ as a function of pressure. This phase is experimentally observed to stabilize between 137 and 218 GPa as a high-temperature
superconductor~ \cite{drozdov2019superconductivity}.
In DFT, LaH${10}$ lies on the hull at 150~GPa. GeoMFormer reproduces this stabilisation pressure, while M3GNet, PET-OAM-XL, and eSEN-30M-OAM stabilise LaH${10}$ somewhat earlier (50--100,GPa).
ORB v3 shows qualitatively different behaviour where LaH${10}$ briefly reaches the hull around 80--100~GPa but destabilises again at higher pressures.

\begin{figure}
    \centering
\includegraphics[width=0.8\linewidth]{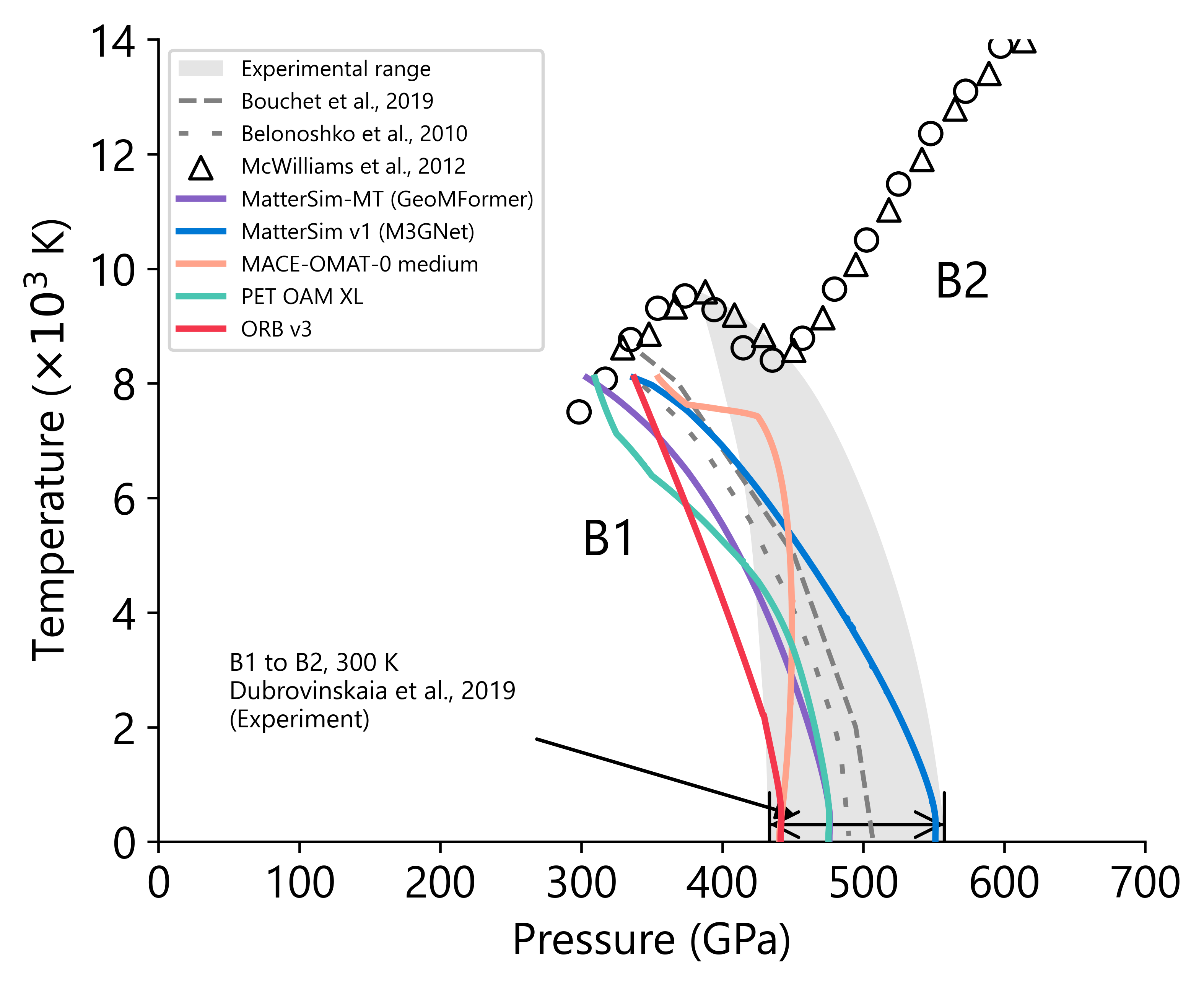}
\caption{B1–B2 phase boundary of MgO computed by DFT and different ML interatomic potentials compared to experimental measurements.}
    \label{fig:case-study-bench-MgO}
\end{figure}

For MgO, we compute the B1–B2 phase boundary using the QHA workflow described in Section\ref{sec:case-study-mgo} and compare to experiments \cite{mcwilliams2012phase,dubrovinskaia2019b1} and \textit{ab-initio} calculations \cite{belonoshko2010mgo,bouchet2019ab} in Fig.\ref{fig:case-study-bench-MgO}.
The shape of the boundary is governed by the Clausius--Clapeyron relation and reflects how the two phases differ in volume and vibrational entropy.
Reproducing this shape therefore tests whether a model captures these thermodynamic quantities correctly.
We note that eSEN-30M-OAM is excluded from this benchmark as the 4$\times$4$\times$4 supercells required for phonon calculations exceeded the memory capacity on a NVIDIA A100 80GB GPU.
MatterSim-MT and MatterSim v1 both reproduce the shape predicted by theory \cite{belonoshko2010mgo, bouchet2019ab}.
While MatterSim v1 overestimates the transition pressure by $\approx$60~ GPa, MatterSim-MT agrees within $\approx$15~GPa at 300~K.
PET-OAM-XL reproduces the boundary at low temperatures but develops an unphysically steep slope above $\approx$5000~K.
ORB v3 produces a nearly linear boundary, failing to capture the curvature.
MACE-OMAT-0 medium yields an essentially flat boundary with no meaningful temperature dependence, indicating that the model does not resolve the free energy difference between the two phases.

\section{MatterSim as a continual active learner}\label{sec:active-learning}

As a universal predictive model for material properties, MatterSim may not yield satisfactory accuracy for highly complex systems that have not been previously encountered in its training dataset. Under such circumstances, an active learning approach can be employed to selectively filter data, followed by finetuning of the MatterSim model. This procedure facilitates the rapid development of a sufficiently accurate and operational model.
In this work, we present an example of the complex ionic superconductor
\ce{Li2B12H12}, as shown in the inset of \autoref{ms-fig:al-and-ft}(a) in the main text. To generate the training dataset, we performed NVT simulations on \ce{Li2B12H12} using VASP\cite{kresse1996efficiency,kresse1996efficient} with a time step of 0.5 fs for a total duration of 5.0 ps at a simulation temperature of 2000 K.
The test dataset belongs to trajectories that are not included in the training set.

The active learning procedure can significantly improve the accuracy of MLFFs by augmenting the dataset with data points that exhibit high uncertainty based on the ensemble models. The uncertainty arises from the variability inherent in different pre-trained models, each initialized with a distinct random number seed. This variability is quantified as follows:
\begin{equation}\label{eq:unc}
\mathrm{unc} = \max \left[ \frac{1}{N} \sum_{i=1}^N  (|F_{ia}| - |\bar{F}_{a}|)^2 \right]
\end{equation}
where $F_{ia}$ denotes the predicted atomic force on the $a$-th atom by the $i$-th model and $N$ represents the total number of pretrained models utilized. $\bar{F}_{a}$ signifies the average of the atomic forces predicted by the ensemble models.

\section{Finetuning and molecular dynamics on liquid water}\label{si-sec:finetuning-and-molecular-dynamics-on-liquid-water}

This section details the application of finetuning to simulate properties of liquid water. We first outline the parameter settings for the finetuning process and MD simulations. Next, we describe the post-processing of the oxygen-oxygen-oxygen angular distribution function (ADF), present the oxygen-hydrogen and hydrogen-hydrogen radial distribution functions (RDF), assess data efficiency, and detail the calculation of diffusion coefficients. The three trained models are referred to as MatterSim-MT, MatterSim scratch-900, and finetune-60. The MatterSim-MT model, trained with PBE-level theory, is discussed in the main text without finetuning and lacks liquid water configurations in its training data. The scratch-900 model is trained from scratch using 900 bulk liquid water configurations from Ref. \cite{cheng2019ab, monserrat2020liquid} with rev-PBE0-D3 level theory. The finetune-60 model is fine-tuned using only 60 of these 900 configurations.

\subsection{Parameter settings for finetuning}\label{sec:finetune-water-setups}
To address the limitations imposed by the level of theory of the training data, we implemented finetuning on the MatterSim model. For finetuning, we discard the pretrained head, reinitialize parameters, and normalize the data. The parameters for finetuning and training from scratch are shown in \autoref{tab:water-finetune-parameters}.

 We randomly sampled 100 liquid water configurations, which were divided into validation and test datasets, with the remaining 900 configurations reserved as potential training candidates. Among the 900 available candidates, 60 configurations were selected at random to create various training sets through the alteration of random seeds, which were subsequently employed to fine-tune the MatterSim model.

\begin{table}[h]
\centering
\caption{Water data training hyperparameters}
\begin{tabular}{lcc}
\toprule
\textbf{Training parameters}  & finetune & from scratch\\
\midrule
optimizer               & AdamW\cite{loshchilov2017decoupled} & AdamW\cite{loshchilov2017decoupled}\\
head max learning rate                     & 2e-4 & 2e-4   \\
backbone max learning rate                     & 5e-5 & 2e-4   \\
weight decay                            & 0.0 & 0.0 \\
$\beta _1$                              & 0.9 & 0.9 \\
$\beta _2$                              & 0.999 & 0.999 \\
$\epsilon$                              & 1e-7 & 1e-7 \\
\midrule
training epochs                              & 200 & 200\\
learning rate schedule                      & stepLR & stepLR \\
step size epochs                            & 8 & 8 \\
\midrule
$\omega_f$                            & 10.0 & 10.0   \\
\bottomrule
\end{tabular}
\label{tab:water-finetune-parameters}
\end{table}

\subsection{Simulation settings for molecular dynamics}
In this work, we evaluated the performance of the finetuning scheme by probing the structural and dynamical properties of liquid water. MD simulations are conducted up to nanoseconds using the ASE package.
The initial structure used for these simulations is composed of a cubic liquid water box containing 512 water molecules with a box length of \SI{24.68}{\AA}.
Production runs of the MD simulations are carried out in the NVT ensemble at \SI{298}{K} and we regulate the temperature using the Nos{\'e}-Hoover thermostat\cite{parrinello1981polymorphic,nose1984unified, martyna1994constant, capinski1997thermal,shinoda2004rapid} every 100 steps. The timestep used to propagate the dynamics is chosen to be \SI{0.5}{fs}. Lastly, the first \SI{200}{ps} out of the nanosecond trajectories are discarded for pre-equilibration.
ADF and RDF, along with diffusion coefficients, are analyzed using the General Purpose Trajectory Analyser (GPTA) software tool.\cite{GPTARepo2023}

\subsection{ADF of Liquid Water}

As illustrated in \autoref{ms-fig:al-and-ft}(d), the ADF of the oxygen species $\left( P_{OOO}(\theta)\right)$, is multiplied by the sine of the angle $\sin (\theta)$, corresponding to the triples of oxygen atoms. The product $P_{OOO}\left(\theta\right)\sin\left(\theta\right)$ allows for a direct comparison with angular distribution extracted from
empirical potential structural refinement (EPSR) based on joint X-ray/neutron scattering measurements,\cite{soper2008quantum} thus enabling better understanding of local arrangements of the water molecules in condensed phase. Following the methodology outlined in previous studies\cite{distasio2014individual, gaiduk2018first, zheng2018structural}, $P_{OOO}(\theta)$ is normalized such that $\int_{0}^{\pi} \ P_{OOO}(\theta) \sin(\theta) \ d\theta$ goes to unity. Similarly, the cutoff value for identifying oxygen triples is chosen so that the oxygen-oxygen coordination number averages around 4.\cite{distasio2014individual}

\subsection{Deriving Diffusion Coefficients from Mean Squared Displacements}\label{si-section:al-and-ft:liquid-water-dissusion}

\begin{figure}
    \centering
    \includegraphics[width=0.75\linewidth]{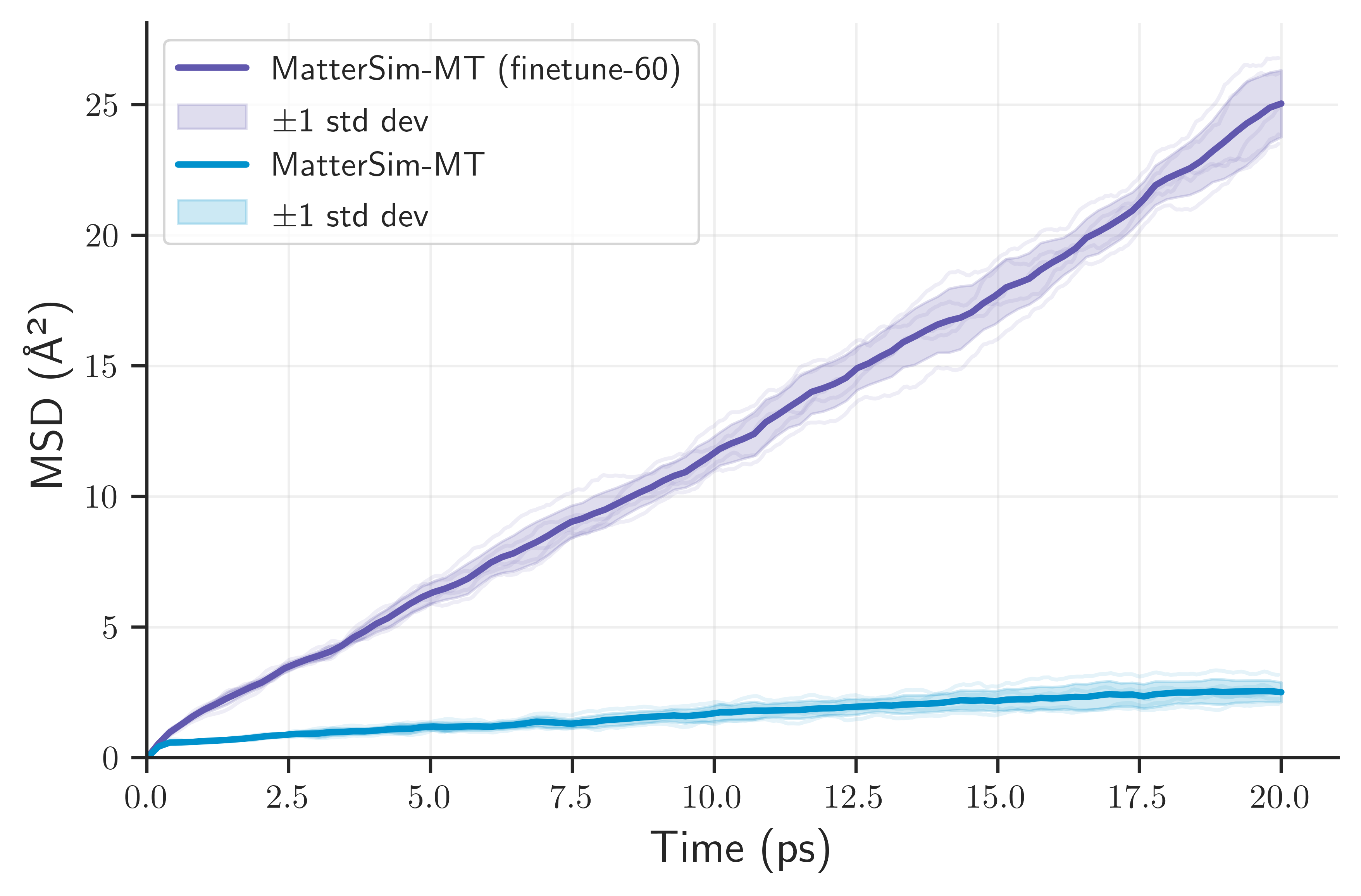}
    \caption{Mean squared displacement (\AA) analysis of MatterSim-MT and finetuned models. Shaded regions indicate one standard deviation of five 20-ps segments from the same trajectory.}
    \label{fig:msd-of-water}
\end{figure}

Diffusion coefficients $\left( \mathcal{D}\right)$ of liquid water at \SI{300}{K} are determined via the Einstein relation\cite{einstein1905annalen, maginn2019best}:

\begin{equation}\label{eq:Einstein_relation}
\mathcal{D} = \frac{1}{6}\lim_{\tau \rightarrow \infty}  \frac{\mathrm{d}\lambda}{\mathrm{d}\tau},
\end{equation}

where $\lambda$ stands for the mean squared displacements along a MD trajectory, and $\tau$ represents the correlation time chosen. \autoref{eq:Einstein_relation} suggests that $\mathcal{D}$ can be obtained by extracting the slope from a linear fit between $\lambda$ and $\tau$. In this work, $\lambda$ is computed along the same trajectory used to compute the RDFs and ADFs. The MSDs computed with the MatterSim-MT and finetuned MatterSim-MT models are reported in \autoref{fig:msd-of-water}, with the shaded regions indicating the standard deviation among 5 segments extracted from the trajectory. As listed in \autoref{tab:Summary_diffusion_coefficients}, MatterSim-MT predicts the diffusion coefficient to be \SI{0.018\pm 0.004}{\angstrom^2\per\pico\second}, which is compatible with simulation results reported in literature, for example \SI{0.018\pm 0.002}{\angstrom^2\per\pico\second} in Ref.~\citenum{chen2017ab}. Moreover, after finetuning with only 60 snapshots of water at the level of rev-PBE0-D3, MatterSim is able to correctly predict the diffusion coefficient to be \SI{0.193\pm 0.011}{\angstrom^2\per\pico\second}, which is also close to the experimental measurements \SI{0.187}{\angstrom^2\per\pico\second},\cite{mills1973self} and \qtyrange{0.230}{0.240}{\angstrom^2\per\pico\second}.\cite{krynicki1978pressure,hardy2001isotope}

\begin{table}[ht!]
    \centering
    \begin{tabular}{c c c }
    \toprule
    Type  & $\mathcal{D} $ (\SI{}{\angstrom^2\per\pico\second}) & Source \\
    \hline
     MatterSim-MT finetune & 0.193$\pm$0.011 & \textbf{This work} \\
     MatterSim-MT & 0.018$\pm$0.004 & \textbf{This work} \\
     Simulation with PBE functional & 0.018$\pm$0.002 & \cite{chen2017ab} \\
     Simulation with SCAN functional & 0.190$\pm$0.025 & \cite{chen2017ab} \\
     Experiment & 0.230$\sim$0.240 & \cite{krynicki1978pressure, hardy2001isotope}\\
     Experiment & 0.187 & \cite{mills1973self}\\
    \bottomrule
    \end{tabular}
    \caption{The diffusion coefficient of water at \SI{300}{\kelvin} predicted with MatterSim-MT and finetuned models of MatterSim-MT, compared with first-principles simulations reported in the literature and experimental measurements.
    }
    \label{tab:Summary_diffusion_coefficients}
\end{table}

\subsection{Data efficiency of finetuning}
\begin{figure}
    \centering
    \includegraphics[width=0.75\linewidth]{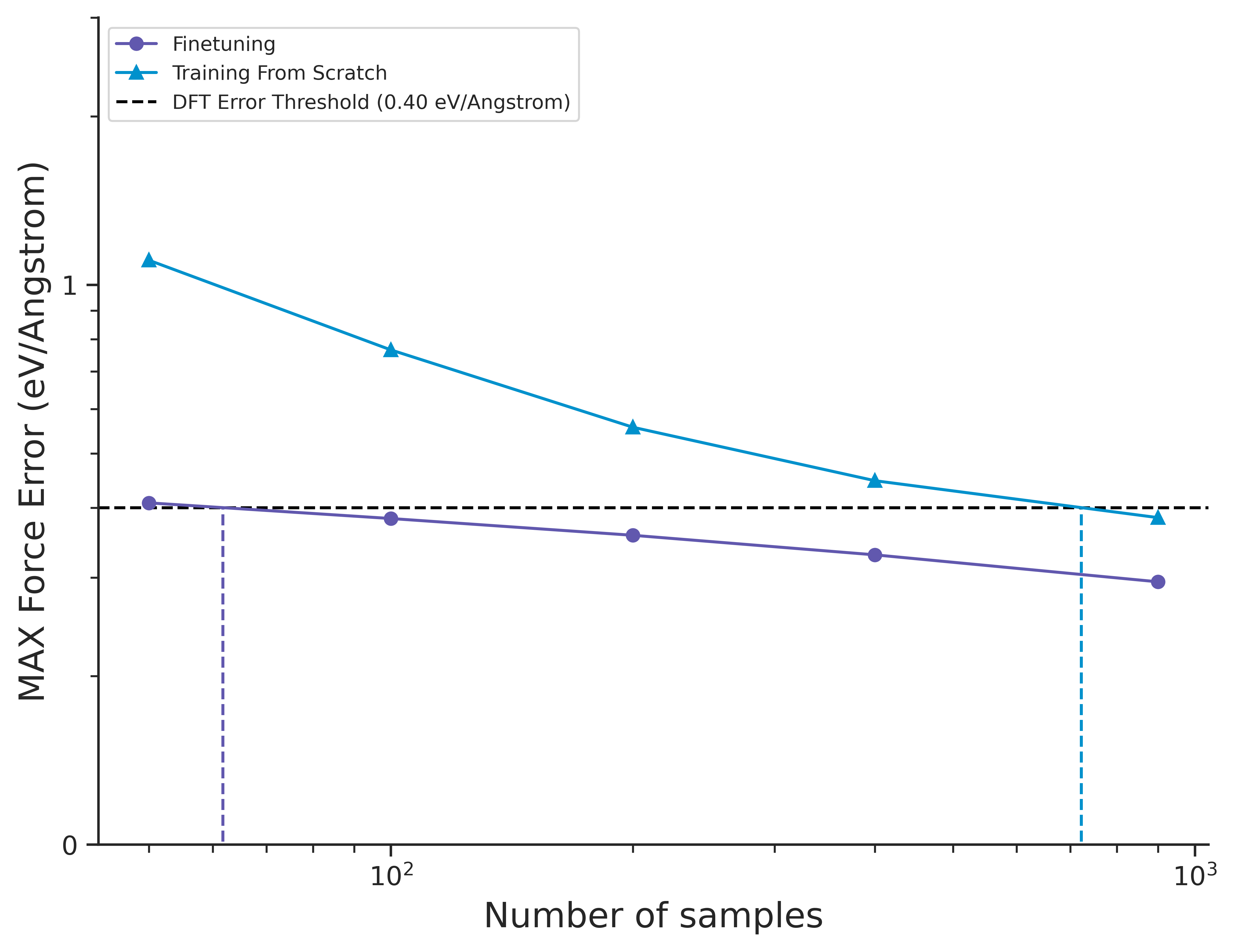}
    \caption{MAX force error against first-principles results in the test set from the finetuned model and the model trained from scratch}
    \label{fig:water-finetune}
\end{figure}

\autoref{fig:water-finetune} illustrates the model's data efficiency during finetuning, showing that only a minimal amount of data is required to achieve the same performance as training from scratch.

\bibliography{ms}